\documentclass[journal]{IEEEtran}

\usepackage{amsmath}
\usepackage{amssymb}
\usepackage{mathrsfs}
\usepackage{bm}

\usepackage{graphicx}
\usepackage{color}
\usepackage{epstopdf}
\usepackage{subfigure} 
\usepackage{subcaption}

\usepackage{booktabs}
\usepackage{multirow}
\usepackage{tabularx}
\usepackage{makecell}

\newcolumntype{C}{>{\centering\arraybackslash}X} 
\usepackage{algorithm}
\usepackage{algorithmic}

\usepackage{cite}
\usepackage{stfloats}
\usepackage{float}
\usepackage{url}
\usepackage{makeidx}
\usepackage[shortlabels]{enumitem}
\usepackage{lipsum} 

\usepackage{caption}
\begin{document}

\title{Stacked Intelligent Metasurface-Diffractive Deep Neural Networks for Onboard Terrain \\ Classification from SAR Level-0 Raw Data
}
\author{\IEEEauthorblockN{Mengbing~Liu, \textit{Graduate Student Member, IEEE},  Xin Li, \textit{Member, IEEE}, \\  Jiancheng An, \textit{Senior Member, IEEE},  and Chau Yuen, \textit{Fellow, IEEE} }\\
 \thanks{Mengbing Liu, Xin Li, Jiancheng An, and Chau Yuen are with the School of Electrical and Electronics Engineering, Nanyang Technological University, Singapore 639798 (e-mails: \tt{mengbing001@e.ntu.edu.sg;  xin019@e.ntu.edu.sg};\tt{jiancheng.an@ntu.edu.sg; chau.yuen@ntu.edu.sg}).}
\thanks{Parts of this paper have been presented at the  International Conference on Learning Representations (ICLR) ML4RS workshop, 2025 \cite{liu2025onboard}.}}

\maketitle
\begin{abstract}

Real-time terrain classification directly from  Level-0 raw Synthetic Aperture Radar (SAR) data remains restricted by traditional digital-centric paradigms, where the processing of noisy, high-dimensional patches is hindered by computationally intensive processors and significant downlink latency. {To address these fundamental limitations, this work establishes a new research paradigm for autonomous on-board sensing by proposing a Stacked Intelligent Metasurface-Diffractive Deep Neural Network (SIM-D$^{2}$NN). This architecture leverages the physical wave-propagation medium to offload inference tasks from digital processors to a physical device. By executing in-wave feature mapping, the SIM-D$^{2}$NN facilitates a move toward an integrated `compute-while-transmitting' framework, providing an alternative to the traditional `digitize-then-process' sequence.} The multi-layer metasurface is positioned at the forefront of the satellite communication module. The initial layer modulates the raw SAR data through both amplitude and phase adjustments, where a 90$^\circ$ phase rotation is introduced as a lightweight but effective augmentation strategy to enhance robustness against noise and Doppler distortions. Subsequent layers learn variable phase shifts, enabling advanced feature mapping for the classification task. The classification results at the terrestrial station can be directly obtained based on the signal amplitude received at each antenna. This design reduces reliance on downlink bandwidth and high-power terrestrial computing, achieving performance around 90\%  in the {binary}  task directly from real raw SAR data in terms of accuracy, precision, recall, and F1 Score.  Therefore, our method helps bridge the gap between next-generation remote sensing tasks and in-orbit processing needs, paving the way for computationally efficient remote sensing applications. 
\end{abstract}


\IEEEpeerreviewmaketitle

\section{Introduction}

\subsection{Background}
On-board remote sensing missions increasingly rely on Synthetic Aperture Radar (SAR) data to support environmental and societal applications such as deforestation detection, flood monitoring, and agricultural assessment \cite{moreira2013tutorial}.  Before the integration of machine learning techniques, the classification of SAR data heavily relied on statistical distributions and physical scattering mechanisms \cite{lee1994classification, cloude1997entropy}. However, these traditional methods of mathematical analysis are limited to extracting only superficial features, which are insufficient for achieving optimal classification accuracy.
\vspace{-5pt}
\subsection{Related Work}
 {
Recent advances in deep learning have significantly enhanced SAR interpretation capabilities, which can be broadly categorized by their primary objectives: land-cover classification, target detection, and high-accuracy image reconstruction \cite{haensch2010complex,zhang2017complex,zhu2017deep, parikh2019classification, ley2018exploiting, liu2024cromss}.} Both \cite{haensch2010complex} and \cite{zhang2017complex} employed complex-valued Convolutional Neural Networks (CNN) for target detection and image classification, respectively. Additionally, more sophisticated models have been applied to SAR data. For example, \cite{ley2018exploiting} shows that classifiers pre-trained on features learned from transcoding SAR to optical images using a conditional generative adversarial network outperform traditional classifiers, especially with limited labeled data. The recent introduction of the cross-modal sample selection method in \cite{liu2024cromss}, which leverages multi-sensor data and noisy labels, has proven effective in enhancing semantic segmentation, as validated by global satellite imagery datasets. {  Despite these advancements, existing digital-centric paradigms decouple data acquisition from semantic inference by treating classification as a post-processing task performed on high-level products like Single Look Complex (SLC) imagery. This fragmented approach necessitates massive raw SAR data transmission, leading to a persistent `digitize-then-process' characterized by significant latency and energy overheads \cite{boser1991analog}. 

To address these limitations, a paradigm shift is unfolding toward processing raw Earth observation data at the source \cite{meoni2024unlocking,del2025enhancing}. Specifically, it is argued that traditional ground-based processing chains are unsuited for onboard applications due to their substantial computational burden and the risk of information degradation, which underscores the critical need for automated raw datasets to foster the development of lightweight, end-to-end edge computing pipelines \cite{meoni2024unlocking}. For instance, Sentinel-2 Level-0 optical imagery for on-board 3-class ship recognition yielded only 50\% per-class accuracy, indicating that raw signals are promising yet difficult to exploit without substantial calibration \cite{del2025enhancing}. 
Similarly, the semantic information in Sentinel-1 (S1) Level-0 raw SAR data is intricately encoded within the deterministic phase history and complex scattering coherence of the targets \cite{moreira2013tutorial}. While recent studies have begun to unlock the optical raw signals for onboard edge computing, the difficulty is more pronounced for the raw SAR data, which remains subject to severe sensor-specific distortions and speckle. Moreover, existing onboard edge computing solutions predominantly rely on digital edge hardware, which encounter significant throughput and energy bottlenecks under strict satellite resource constraints. This motivates a strategic shift from energy-intensive digital processors toward an in-wave analog computing approach that performs feature extraction during the signal propagation process itself.
 
  By treating the electromagnetic environment as a natural computational medium, in-wave processing enables a `compute-while-transmitting' framework \cite{lin2018all, yan2019fourier,  mengu2019analysis, liu2022programmable,ma2026diffractive,liu2026in,niu2026two}.
This perspective shifts the research focus from digital post-processing to in-wave manipulation, where an optical Diffractive Deep Neural Network  (D$^2$NN) architecture was introduced to perform in-wave execution of tasks \cite{lin2018all}.} {
To further augment the representational capacity of such physical pipelines, a Fourier-space D$^2$NN framework was developed in \cite{yan2019fourier} to natively execute high-frequency diffractive features within the spatial-frequency spectrum.}
  Building on this foundation, the authors presented further advancements in analog neural networks, enhancing performance through improved training techniques and integrating electronic neural networks to develop a hybrid optical-electronic neural network  \cite{mengu2019analysis}, which effectively reduced input size and complexity while maintaining high performance.
Transcending passive designs, active D$^2$NNs utilizing programmable multilayer metasurfaces with amplifier-integrated meta-atoms have been proposed to offload computation directly into the wave domain \cite{liu2022programmable}.  

{
Similarly, in the wireless communication domain, Reconfigurable Intelligent Surfaces (RIS) have transitioned from pure signal enhancement to sophisticated information processing.}
While a substantial body of work has addressed fundamental communication challenges, such as channel estimation \cite{wei2021channel, liu2022deep, TCOM_2022_An_Low} and beamforming \cite{huang2021multi, Samith2020Intelligent,liu2025beamforming}, recent efforts leverage RIS to dynamically reshape the wireless environment for computational tasks \cite{sanchez2022airnn, chen2024RIS,liu2025over}.
In detail, the authors of \cite{sanchez2022airnn} explored the design of an analog CNN by utilizing reconfigurable intelligent surfaces (RISs) to shape wireless propagation, enabling inference directly within the analog domain for modulation classification. Similarly, an RIS-based semantic communication paradigm was described that leverages a D$^2$NN  to facilitate in-wave computation, achieving high-speed semantic transmission with minimal power consumption \cite{chen2024RIS}.
Recently, an over-the-air ordinary differential equation-inspired neural network was proposed to utilize distributed  RISs as an in-wave 
block for feature mapping in image reconstruction \cite{liu2025over}.

{Building on these successes on RIS, Stacked Intelligent Metasurfaces (SIMs) have emerged as a high degree-of-freedom solution for their ability to enhance system efficiency and performance through multiple layers of programmable phase control \cite{TAP_2024_An_Emerging,yang2024joint,WC_2024_An_Stacked,sun2025dual,Lin2025UAV,Papaz2025performance,an2023stacked, li2025stacked, shi2025uplink, huang2024stacked, An2024two}.} For instance,  SIMs are integrated into wideband holographic Multiple-Input Multiple-Output (MIMO) systems to optimize phase shifts and digital precoding, improving spectral efficiency and mitigating phase tuning errors \cite{li2025stacked}. Meanwhile, \cite{shi2025uplink} explores an SIM-assisted cell-free massive MIMO framework, introducing beamforming and power control strategies, achieving a 57\% spectral efficiency gain over traditional cell-free massive MIMO. Studies such as \cite{huang2024stacked} and \cite{An2024two} have demonstrated the practical deployment of SIMs, enabling advanced functionalities like image recognition and sensing in transceiver communications.

\subsection{Technical Contribution}

Given the potential of SIMs for programmable in-wave analog computation, we propose an SIM-D$^2$NN architecture for onboard terrain classification. This system processes  S1 raw SAR data directly, minimizing terrestrial data transmission and enabling real-time, low-power inference. However, raw SAR data lacks structured features, and the SIM’s linear nature complicates optimization for complex tasks. To address these challenges, our contributions are:
 
\begin{itemize} 
 
{\item 
\textbf{`Compute-While-Transmitting' Framework}: We propose the SIM-D$^2$NN framework to establish a new paradigm for on-board sensing. By utilizing the multi-layer SIM as a physical computational medium, this architecture facilitates a transition from the traditional `digitize-then-process' sequence to a unified `compute-while-transmitting' framework. This design enables autonomous, progressive linear feature mapping directly within the electromagnetic domain during signal propagation, thereby offloading inference tasks from energy-intensive digital processors. 

 \item \textbf{In-wave Augmentation for Raw SAR Data Learnability}: 
To address the limited learnability of noisy raw SAR data, stemming from inherent speckle and Doppler distortions, we introduce a lightweight $90^{\circ}$ phase-rotation strategy. Validated through comprehensive ablation studies, this approach serves as a computationally efficient physical-layer alternative to traditional digital preprocessing pipelines. By performing this minimal adjustment during in-wave manipulation, we successfully unlock the learning potential of raw SAR data, elevating {binary}  task accuracy from a random-guess baseline of 50\% to 90\%, indicating that reliance on certain digital preprocessing pipelines can be mitigated through strategic phase modulation within the physical in-wave medium.
 
\item \textbf{Performance Evaluation and the Limits of Linearity:} 
We provide a comprehensive assessment of the linear representational capacity of multi-layer diffractive architectures by extending the evaluation to multi-class recognition tasks across diverse regions. While the multi-layer design yields a 26\% accuracy improvement over single-layer counterparts, we identify clear operational boundaries for purely linear analog processing when handling raw SAR data, where accuracy declines from 89.5\% in { binary}  tasks to 63.2\% in complex  {four}-class scenarios. To provide a baseline for the architecture's inherent capability, we benchmark the system against a 14-class high-level SAR dataset, where it attains an accuracy of 93\%. This high performance confirms that linear in-wave operations remain highly effective when the data have been thoroughly pre-processed, yet also underscores the necessity of integrating nonlinear mechanisms for future architectures aimed at tackling complex semantic tasks directly from raw SAR data.}
\end{itemize}

\subsection{Paper Organization}

The remainder of this paper is organized as follows. Section II introduces the system model for the  SIM-D$^2$NN framework, describing the space-to-ground channel and the labeling procedure for raw SAR data patches. Section III explains the problem formulation and presents the training and deployment algorithm used to optimize metasurface phase shifts for effective feature mapping. Section IV offers a comprehensive evaluation of how different network configurations and training strategies affect the terrain-classification performance of the proposed SIM-D$^2$NN. Section V reports simulation results on raw SAR data and demonstrates the generalizability of the SIM-D$^2$NN on high-level SAR products. Finally, Section VI concludes the paper and outlines avenues for future research.

\textit{Notation:} Bold uppercase and lowercase letters denote matrices and vectors, respectively. The sets of real and complex numbers are represented by $\mathbb{R}$ and $\mathbb{C}$. $(\cdot)^T$ and $(\cdot)^H$ denote the transpose and Hermitian transpose operations, while $\operatorname{diag}(\cdot)$ represents a diagonal matrix. $\|\cdot\|_{\mathcal{F}}$ denotes the Frobenius norm. For a complex value $z$, $j=\sqrt{-1}$ is the imaginary unit, and $\operatorname{Re}\{z\}$, $\operatorname{Im}\{z\}$, and $|z|$ denote its real part, imaginary part, and modulus, respectively. $\mathbb{I}{(\cdot)}$ is the indicator function.  $v \sim \mathcal{CN}(0,\sigma^2)$ represents a circularly symmetric complex Gaussian distribution with zero mean and variance $\sigma^2$.

\section{SIM-aided Satellite Communication System}
This section details the SIM-aided satellite system for onboard terrain classification. We present the system model, space-to-ground channel characteristics, and the labeling methodology for the  S1 Level-0 raw SAR dataset.
\subsection{System Model}
\begin{figure}[th]
 \begin{center}
    \includegraphics[width=  1\linewidth]{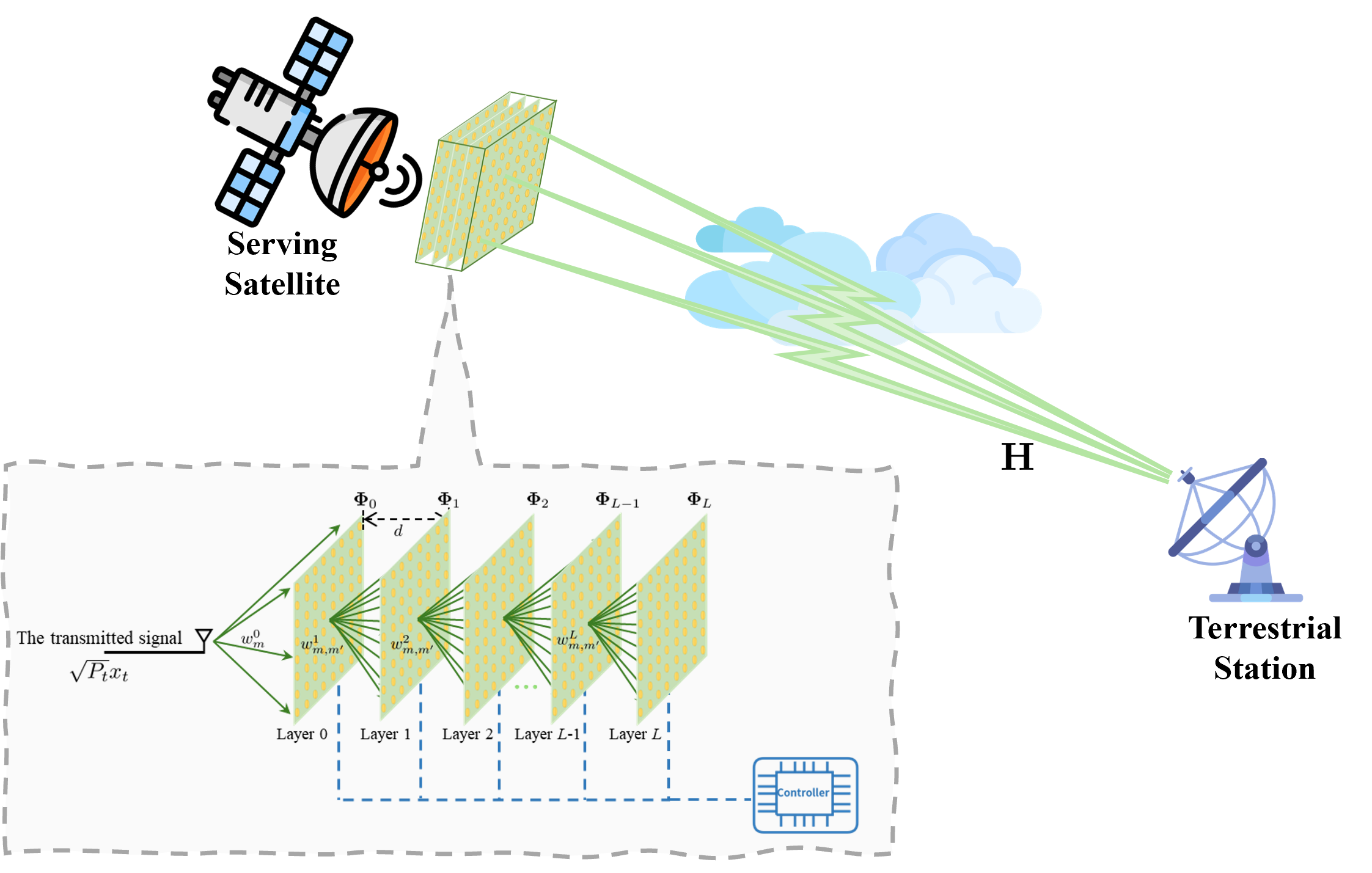}
 \end{center}
    \caption{The system model with SIM-D$^2$NN structure.}
    \label{system_model}
\end{figure}

First, as illustrated in Fig.~\ref{system_model}, we explore an SIM-aided wireless system comprising a single-antenna satellite and a $K$-antenna terrestrial station.
The SIM consists of \(L+1\) metasurface layers, denoted by the set \(\mathcal{L} = \{0, 1, 2, \dots, L\}\), with each layer containing \(M\) meta-atoms, represented by the set \(\mathcal{M} = \{1, 2, \dots, M\}\). For each metasurface layer in the SIM,  the diagonal phase shift matrix for the \(l\)-th layer is expressed as \(\mathbf{\Phi}^{l} = \text{diag}(a_{1}^{l}e^{j\theta_{1}^{l}}, a_{2}^{l}e^{j\theta_{2}^{l}}, \dots, a_{M}^{l}e^{j\theta_{M}^{l}}) \in \mathbb{C}^{M \times M}, \forall l \in \mathcal{L}\), where \(\theta_{m}^{l} \in [0, 2\pi)\) and \(a_{m}^{l}\) represents the phase shift and magnitude of the \(m\)-th meta-atom in the \(l\)-th layer, respectively.  
Here, the meta-atoms of the 0-th layer $\mathbf{\Phi}_i^0 $  are reconfigured using a controller to align with the augmented input features of the $i$-th patch $\boldsymbol{\overline{s}}_{i}$, {acting as a complex-valued spatial modulator}, enabling efficient manipulation of electromagnetic waves in the wave domain. Thus, the value deployed at the $0$-th layer for the $i$-th patch is given as 
\begin{align}
    \mathbf{\Phi}_i^0 
&= 
\operatorname{diag}\left( \overline{s}_{i,1},\, \overline{s}_{i,2},\, \dots,\, \overline{s}_{i,M} \right),
\end{align}
where $\overline{{s}}_{i,m}$ is the $m$-th element in the normalized input tensor after data augmentation. The subsequent layers $\mathbf{\Phi}^{l} = \text{diag}(e^{j\theta_{1}^{l}}, e^{j\theta_{2}^{l}}, \dots, e^{j\theta_{M}^{l}}) \in \mathbb{C}^{M \times M},   l \in \{1,2,\ldots, L\}$ apply learned phase shifts $\theta_{m}^l$ to the incoming wave to enable deep feature mapping. {Mathematically, these layers represent a sequence of programmable diagonal operators that reconfigure the phase of each meta-atom $\theta_{m}^l$  based on the output for optimizing performance \cite{lin2018all}.}

Let \(\mathbf{W}^{l} \in \mathbb{C}^{M \times M}, \forall l \neq 0, l \in \mathcal{L}\), denote the transmission matrix between the \((l-1)\)-th and the \(l\)-th metasurface layers. {  Based on the Rayleigh-Sommerfeld diffraction theory \cite{lin2018all}, the transmission matrix $\mathbf{W}^{l}$ physically implements full complex connectivity between successive layers, where each meta-atom acts as a secondary Huygens source. 
}
The \((m, m')\)-th element \(w_{m,m'}^{l}\) of the matrix \(\mathbf{W}^{l}\) is described as
\begin{equation}
w_{m,m'}^{l} = \frac{d_x d_y \cos \chi_{m,m'}}{d_{m,m'}^{l}} \left( \frac{1}{2\pi d_{m,m'}^{l}} - j\frac{1}{\lambda} \right) e^{j2\pi d_{m,m'}^{l}/\lambda},
\label{w}
\end{equation}
where \(\lambda\) denotes the wavelength, \(d_{m,m'}^{l}\) is the transmission distance between the meta-atoms, \(\chi_{m,m'}\) is the angle between the propagation direction and the normal of the \((l-1)\)-th metasurface layer, and \(d_x \times d_y\) represents the dimensions of each meta-atom. Similarly, the \(m\)-th entry \(w_{m}^{0}\) of \(\mathbf{w}^{0}\)  can be derived from (\ref{w}).
 Thus, the propagation matrix \(\mathbf{G}_i \in \mathbb{C}^{M \times M}\) for the $i$-th  { patch {can be interpreted as a reconfigurable spatial matched filter}. Within this framework, the cascaded metasurface layers physically synthesize a complex spatial impulse response to perform in-wave correlation between the raw SAR echoes and the learned terrain spatial signatures as}
\begin{equation}
\mathbf{G}_i = \mathbf{\Phi}^L\mathbf{W}^L\mathbf{\Phi}^{L-1} \cdots \mathbf{\Phi}^{1}\mathbf{W}^{1}\mathbf{\Phi}^{0}_i.
\label{sim}
\end{equation}

Under this propagation model, { the SIM-D$^2$NN physically executes high-dimensional cascaded matrix-vector multiplications in the wave domain, performing spatial spectrum reconfiguration at the speed of light \cite{lin2018all, liu2022programmable}. While the wave propagation remains linear, the multi-layer stacking boosts the spatial degrees of freedom, yielding a matched-filtering response for terrain feature extraction, bypassing the need for computationally intensive digital signal processing.}
 Moreover, the SIM's parallel nature ensures that processing speed is independent of the number of meta-atoms per layer.
  \begin{figure*}[t]
\begin{center}
    \includegraphics[width= 1\textwidth]{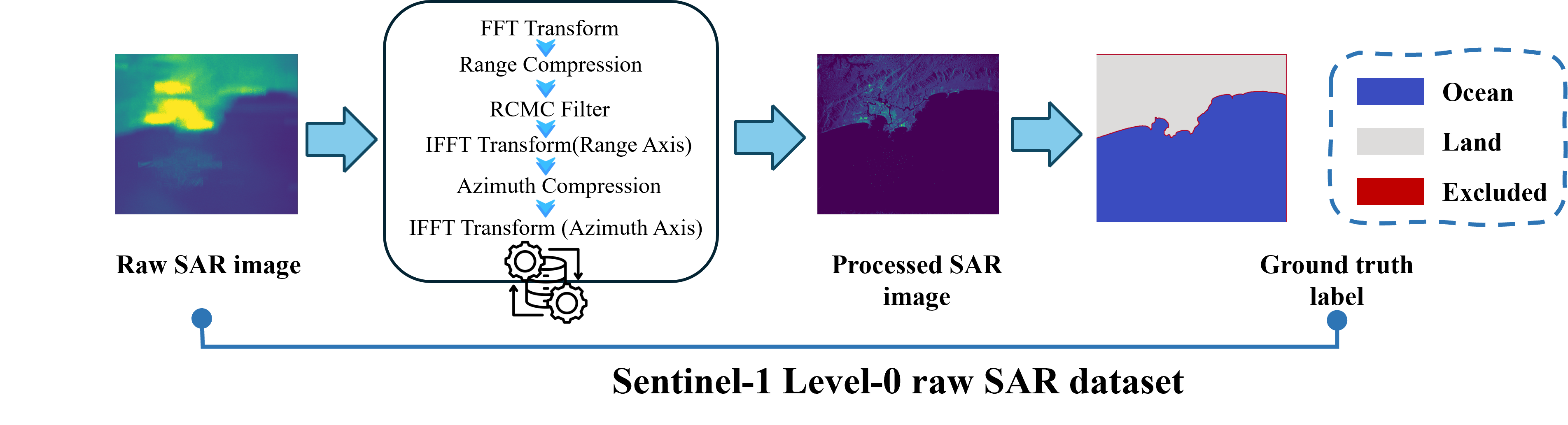}
\end{center}
\caption{ \textbf{The workflow for labeling the raw SAR data.} The workflow for labeling raw SAR data is illustrated using SAR data collected over Sao Paulo and the nearby port of Santos, Brazil \cite{hall2023sentinel1level0decoding}. This visualized region serves as a basis for sampling patches of land and ocean for terrain classification. Based on the processed image, the region is segmented into ocean (blue), land (grey), and excluded areas (red). The excluded areas represent boundaries unsuitable for labeling due to mixed terrains. }
    \label{workflow}
\end{figure*}

Furthermore, let $\sqrt{P_t}x_t$ represent the transmitted signal, with $P_t$ being the transmit power and $x_t$ the normalized electromagnetic signal such that $|x_t|^2 = 1$. 
As the signal $\sqrt{P_t}x_t$ propagates through the initial layer, it carries the encoded input data to the subsequent layers, where feature mapping and onboard classification are automatically performed. This setup requires receiving merely the classification outcomes with a small amount of data, thereby streamlining the overall data processing workflow. 
 Then, the received signal $\mathbf{y}_i \in \mathbb{C}^{K \times 1}$ for the $i$-th patch is given as
\begin{equation}
    \mathbf{y}_i = \mathbf{H}\,\mathbf{G}_i\,\mathbf{w}^0\,\sqrt{P_t}x_t + \mathbf{n},
\end{equation}
where  $\mathbf{H} \in \mathbb{C}^{K \times M}$ and  \(\mathbf{n} \sim \mathcal{CN}(0, \sigma^2)\) represent the channel from the SIM to the terrestrial station and the i.i.d. additive white Gaussian noise with noise power \(\sigma^2\). 

 After receiving $\mathbf{y}_i$ from $K$ antennas, classification is performed by selecting the antenna with the highest signal magnitude\footnote{Unlike traditional array-aided systems, the received signal $\mathbf{y}_i$ in SIM-D$^2$NN undergoes a controlled analog transformation. By carefully tuning the phase shifts at each metasurface layer $\mathbf{\Phi}_l$, the receiver antenna array can directly extract the likelihood of the incident signal’s category, significantly streamlining hardware design and simplifying subsequent signal processing. This approach also contributes to lower energy consumption by reducing computational overhead.}. This process is mathematically expressed as   
\begin{equation}
     \hat{k}_i
= \arg\max_{k \in \mathcal{K}}
\left\{
|y_{i,1}|^2,\,
|y_{i,2}|^2,\dots,
|y_{i,K}|^2
\right\}, 
\end{equation}
{
where $\hat{k}_i$ denotes the index of the receive antenna capturing the maximum signal intensity for patch $i$, and $\mathcal{K} = \{1, 2, \dots, K\}$ represents the set of all receiving antennas.}
  { While wave propagation is linear, the square-law intensity detection serves as a physical-domain activation \cite{liu2022programmable}, {consistent with this matched-filter interpretation}, enabling the SIM-D$^2$NN to establish discriminative decision boundaries for terrain classification.}
\subsection{Space-to-Ground Channel Modeling}
For the wireless link, unlike terrestrial communication scenarios, where channels in urban areas are typically modeled as Rayleigh fading, the space-to-ground channel is modeled using a Rician fading model, which accounts for both Line-of-Sight (LoS) and Non-Line-of-Sight (NLoS) components with Rician factor   \cite{paulraj2003introduction},  the channel  matrix \(\mathbf{H}\) is expressed as 
\begin{equation}
 \mathbf{H} = \sqrt{{\rm PL}(d, f)} \left(\sqrt{\frac{\kappa}{1 + \kappa}} \mathbf{H}_{\text{LoS}} + \sqrt{\frac{1}{1 + \kappa}} \mathbf{H}_{\text{NLoS}} \right),
\end{equation}
where \(\mathbf{H}_{\text{LoS}}\) and \(\mathbf{H}_{\text{NLoS}}\) represent the channel matrices for the LoS and NLoS components, respectively, and \(\kappa\) is the Rician factor. Additionally, the path loss from the SIM to the terrestrial station, i.e., the satellite-to-ground path loss, is modeled as  \cite{al2020modeling}
\begin{equation}
    {\rm{PL}}(d, f) = {\rm{FSPL}}(d, f) + {\mathscr{L}_{A}}(f,\psi)  + {\mathscr{L}_{E}}(\psi),
\end{equation}
where \(f\) and \( d\) is the carrier frequency and the distance, respectively. \( {\mathscr{L}_{A}}(f,\psi) \) represents the attenuation due to atmospheric absorption, and \({\mathscr{L}_{E}}(\psi)\) characterizes the path loss due to interactions with near-surface urban structures. Then, the expression of  the free space path loss is given as
\begin{equation}
    {\rm{FSPL}}(d, f)= 20\log(f) + 20\log(d) - 147.55.
\end{equation}

 \subsection{Raw SAR Data Labeling}

Contrary to previous studies that predominantly utilize high-level SAR products for terrain classification, our research is dedicated to directly exploiting raw SAR data. Typically, datasets are comprised of SLC and other advanced data formats. Our goal is to establish the practical feasibility of the SIM-D$^2$NN framework for near real-time classification within satellite operations, necessitating the creation of a dataset that encompasses raw SAR data alongside their corresponding ground truth terrain labels.
 
 Directly visualizing raw SAR data is hindered by inherent noise and Doppler distortions, which obscure terrain features and impede reliable ground-truth labeling. Unlike high-level SAR products, raw SAR data preserve intact phase and amplitude information but lack radiometric and geometric corrections. To generate accurate reference imagery for labeling, we utilize an open-source   S1 Level-0 decoding algorithm \cite{hall2023sentinel1level0decoding}. The processing pipeline, illustrated in Fig.~\ref{workflow}, transforms raw SAR data  into the frequency domain via a Fast Fourier Transform (FFT) to execute the following core steps 

\begin{itemize}
 
   \item \textbf{Range compression.}  
          A matched filter enhances range resolution, yielding
          \begin{equation}
              \mathscr{F}_{\text{range}}
              = \exp\!\Bigl\{\,2\pi j\bigl[\bigl(f_{\text{start}}+\tfrac{\alpha T_p}{2}\bigr)\tau
              + \tfrac{\alpha}{2}\tau^{2}\bigr]\Bigr\},
              \label{eq:range_comp}
          \end{equation}
          where $f_{\text{start}}$ is the chirp start frequency, $\alpha$ the chirp rate, $T_p$ the pulse duration, and $\tau$ the fast‐time variable.

   \begin{figure*}[t]
\begin{center}
    \includegraphics[width=0.88\textwidth]{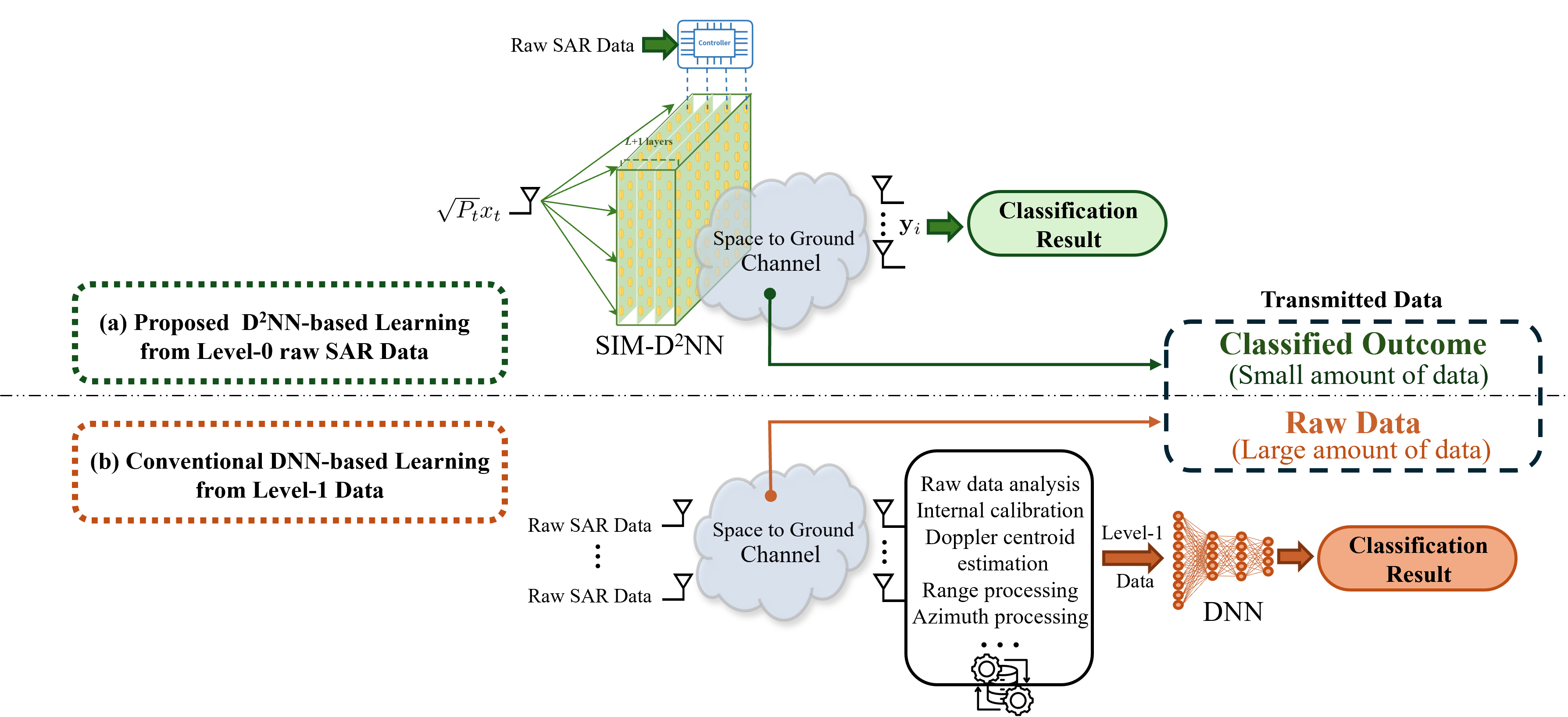}
\end{center}
\caption{
    Traditional digital-domain inference on Level-1 products versus the proposed SIM-D$^2$NN  in-wave inference on Level-0 raw SAR data, illustrating the transition to an integrated `compute-while-transmitting' architecture.}
    \label{system}
\end{figure*}
 \item \textbf{Range‐cell migration correction (RCMC).}  
          Motion‐induced range walk is compensated by
          \begin{equation}
              \mathscr{F}_{\text{RCMC}}
              = \exp\!\Bigl\{\,4\pi j\,f_{\tau} R_{0}\!\bigl(D^{-1}-1\bigr)\!/c\Bigr\},
              \label{eq:rcmc}
          \end{equation}
          where $R_{0}$ is the slant range to the scene centre, $D$ the cosine of the squint angle, $f_{\tau}$ the range frequency, and $c$ the speed of light.
          
  \item \textbf{Azimuth compression.}  
          After an Inverse FFT (IFFT) returns the data to the range domain, azimuth focusing is carried out with
          \begin{equation}
              \mathscr{F}_{\text{azimuth}}
              = \exp\!\Bigl\{-\,4\pi j\,R_{0} D / \lambda\Bigr\},
              \label{eq:az_comp}
          \end{equation}
          where $\lambda$ denotes the radar wavelength.
\end{itemize}
 Finally, an IFFT along each azimuth line converts the focused data to the spatial domain. The right panel of Fig.~\ref{workflow} is the resulting imagery, whose reduced noise and enhanced clarity enable precise terrain labeling for subsequent training.

{
  This section establishes the physical parameters of the SIM-D$^2$NN and the specific methodology for constructing the raw SAR dataset. These characterizations provide the necessary basis for the formal formulation of the terrain classification objective and its associated optimization algorithm.}
\section{Problem Formulation and Proposed SIM-D$^2$NN Framework}
This section formulates classification as a metasurface phase optimization problem, contrasting the SIM-D$^2$NN pipeline with digital workflows across all layers.
 As depicted in Fig.~\ref{system}(a), {the proposed `compute-while-transmitting' framework maps the Level-0 raw SAR data directly onto the initial metasurface layer, where the input comprises unprocessed and unfocused complex radar echoes prior to conventional SAR focusing and Level-1 image reconstruction.} As the signal propagates through successive layers of the SIM-D$^2$NN, automatic feature mapping and onboard terrain classification are performed. This design transmits only final semantic classification results rather than high-dimensional raw SAR data to the terrestrial station, effectively bypassing the computational overhead and latency associated with buffering and electronic reconstruction in digital processors.
In contrast, { the conventional `digitize-then-process' pipeline} illustrated in Fig.~\ref{system}(b) relies on transmitting large volumes of raw SAR data to the ground station,  where digital processing pipelines are carried out { to obtain the Level-1 imagery products for further deep learning inference}. Due to the absence of high-performance computing capabilities onboard, such methods impose significant latency and communication burdens. Thus, the proposed SIM-D$^2$NN architecture outperforms the conventional DNN-based pipeline in terms of onboard efficiency, latency, and communication overhead.

 \subsection{Problem Formulation}
Given the raw SAR dataset $\mathcal{S} = \{({\boldsymbol{S}_{i}}, \boldsymbol{o}_i)\}_{i = 1}^{I}$, where ${\boldsymbol{S}_{i}}$ denotes the $i$-th patch and  $\boldsymbol{o}_i \in \mathbb{R}^{C \times 1}$ is the corresponding one-hot encoded ground truth label vector with $C$ categories. The objective is to optimize the phase shifts in the SIM to emulate a D$^2$NN and maximize classification accuracy via minimizing the discrepancy between the network outputs and the ground truth. Therefore, the optimization problem is given as 
\begin{subequations}
\begin{align}
\text{P1: } \min_{ \substack{ \{\theta_{m}^{l}\} \\ \forall m \in \mathcal{M}, \forall l \in \mathcal{L} } } \quad & \mathcal{L} (|\boldsymbol{y}_{i}|, \boldsymbol{o}_i)  \\
\text{s.t.} \quad & \mathbf{G}_i = \mathbf{\Phi}^L \mathbf{W}^L \mathbf{\Phi}^{L-1} \cdots \mathbf{\Phi}^{1} \mathbf{W}^{1} \mathbf{\Phi}^{0}_i, \\
& \mathbf{\Phi}_i^0 = \mathrm{diag} \left( \overline{s}_{i,1}, \overline{s}_{i,2}, \ldots, \overline{s}_{i,M} \right), \\
& \mathbf{\Phi}_l = \mathrm{diag}(e^{j\theta_{1}^{l}}, e^{j\theta_{2}^{l}}, \dots, e^{j\theta_{M}^{l}}), \\
& |e^{j\theta^{l}_{m}}| = 1, \quad \forall m \in \mathcal{M}, \forall l \in \mathcal{L},
\end{align}
\label{optimization}
\end{subequations}
\noindent where Eq.~(12a) states the optimization objective to minimize the  Cross-Entropy (CE) loss by adjusting the learnable parameters \(\theta_{m}^{l}\) for each layer \(l\) with \(l = 1, \ldots, L\). 
The constraints in Eqs.~(12b)$\sim$(12d) characterize the physical signal transmission {as a cascaded complex-valued linear operator}, while Eq.~(12e) imposes the unit-modulus requirement on each meta-atom. Due to the intricate inter-layer coupling within P1, solving the optimization problem is non-trivial. Consequently, the following subsection details a specialized training and deployment algorithm for the SIM-D$^2$NN framework to facilitate onboard terrain classification. {Specifically, consistent with physical intensity detection, the optimization minimizes the CE loss between the received signal modulus $|\mathbf{y}_i|$ and label $\boldsymbol{o}_i$ as}
  \begin{align}
      \mathcal{L} (|\boldsymbol{y}_{i}|,\boldsymbol{o}_i ) = -\sum_{{c}=1}^{C}  {o}_{i,{c}} \log\left( \frac{e^{| {y}_{i,k=c} |}} {\sum_{k' = 1}^{K} e^{ |{y}_{i,k'} |}}\right).
      \label{loss}
 \end{align}
{To update the phase shifts during the offline training stage, the gradient of the real-valued loss $\mathcal{L}$ with respect to the real parameter $\theta_m^l$ is derived via the {complex-valued chain rule applied to the diffractive field \cite{lin2018all}}. Let $\mathbf{u}_i^l \in \mathbb{C}^{M \times 1}$ represent the intermediate complex field vector after the $l$-th metasurface layer, which can be expressed based on \eqref{sim} as $\mathbf{u}_i^l = \mathbf{\Phi}^l \mathbf{W}^l \dots \mathbf{\Phi}_i^0 $. The forward propagation is thus a differentiable function of  phase shifts, yielding
\begin{equation}
\frac{\partial \mathcal{L}}{\partial \theta_m^{l}} = \text{Re} \left\{ \frac{\partial \mathcal{L}}{\partial u_{i,m}^{l}} \cdot \frac{\partial u_{i,m}^{l}}{\partial \theta_{m}^{l}} \right\} = \text{Re} \left\{ \frac{\partial \mathcal{L}}{\partial u_{i,m}^{l}} \cdot (j u_{i,m}^{l}) \right\},
\end{equation}
where $u_{i,m}^{l} = e^{j\theta_m^l} v_{i,m}^{l}$ is the $m$-th component of $\mathbf{u}_i^l$, and $v_{i,m}^{l}$ is the incident field component arriving at the $m$-th meta-atom from the preceding layer. {
The optimization process thus directly tunes the physical phase configurations to approximate a task-specific matched-filtering response \cite{lin2018all,liu2022programmable}.}
}

 \subsection{Procedures of the SIM-D$^2$NN}
 
This subsection first presents a phase-rotation augmentation that enriches the statistical diversity of raw SAR data patches, then details the two-stage workflow by which the SIM-D$^2$NN performs onboard terrain classification.

\subsubsection{Phase-rotation Augmentation}
                            {
Before the phase-rotation augmentation, each $128\times128$ complex Level-0 raw SAR patch is first magnitude-normalized to $[0,1]$  while preserving its phase. Specifically, letting $\widetilde{\boldsymbol{S}}_i \in \mathbb{C}^{128 \times 128}$ denote the magnitude-normalized Level-0 raw patch, the non-overlapping $4\times4$ coherent spatial averaging is given by
\begin{equation}
\boldsymbol{S}_i^{(p,q)}=\frac{1}{16}
\sum_{u=0}^{3}\sum_{v=0}^{3}
\widetilde{\boldsymbol{S}}_i^{(4p+u,\,4q+v)},
\end{equation}
yielding a $32\times32$ spatially aggregated complex input representation $ \boldsymbol{S}_i$. This coherent spatial averaging provides a discrete representation of localized aperture aggregation over non-overlapping pixel-equivalent regions.}
{
To enrich the representation of the phase information in the complex-valued Level-0 SAR data, we introduce a phase-domain feature augmentation by applying a fixed $90^\circ$ phase rotation to the spatially aggregated complex input coefficients as
}
 {\color{black}\begin{algorithm}[htb]
\caption{Training and Deployment of the SIM-D$^2$NN}
\label{alg:Framwork}
\begin{algorithmic}[1]
\REQUIRE ~~\\
Sliding window patches $\mathcal{S} = \{({\boldsymbol{S}_{i}, \boldsymbol{o}_i})\}_{i = 1}^{I}$;\\
Channel matrix $\mathbf{H}$; fixed weights $\mathbf{W}^{l}$ for $l \neq 0$ and $\mathbf{w}^{0}$.
\ENSURE ~~\\
Terrain classification results on S1 raw SAR data.

\STATE \textbf{Stage 1: Offline Training}
\STATE {Sample a subset patches from the isolated training set.}
\FOR{epoch $= 1$ to $N_{e}$}
    \STATE Apply modulation to the input: \( \overline{\mathbf{s}}_{i} \gets \text{Modulation}( {\mathbf{S}}_{i}) \);
    \STATE Initialize the input metasurface: \\ \( \mathbf{\Phi}_i^0 \gets \text{diag}( \overline{s}_{i, 1}, \overline{s}_{i, 2}, \dots, \overline{s}_{i, M}) \);
    \STATE Compute output $\mathbf{y}_{i}$ via Eq.~(4) and update phase shifts ${\mathbf{\theta}}_m^{l}$ using backpropagation.
\ENDFOR
\RETURN Optimized  phase shifts $\hat{\mathbf{\theta}}_m^{l}$, $\forall m \in \mathcal{M}, l \neq 0, l \in \mathcal{L}$.

\STATE \textbf{Stage 2: Online Deployment}
\FOR{each patch $i = 1$ to $I$}
    \STATE Apply the same modulation procedure to obtain $\overline{\mathbf{s}}_{i}$;
    \STATE Configure each metasurface layer using $\hat{\mathbf{\theta}}_m^{l}$;
    \STATE Compute the output $\mathbf{y}_i$ according to Eq.~(4);
    \STATE Predict label by: \\
        $ \hat{k}_i = \arg\max\limits_{k \in \mathcal{K}} \left\{|y_{i,1}|^2,\, |y_{i,2}|^2,\dots, |y_{i,K}|^2 \right\},$ 
\ENDFOR
\RETURN Final classification results for all patches.
\end{algorithmic}
\end{algorithm}}
 \begin{equation} \boldsymbol{\hat{S}}_i^{(p,q)} = e^{j {\theta}_{\rm rot}} \cdot \boldsymbol{S}_i^{(p,q)}, \end{equation} 
where $\theta_{\rm rot}$ denotes the rotation angle and $\boldsymbol{\hat{S}}_i^{(p,q)}$ represents the $(p,q)$-th complex-valued {coefficient of the spatially aggregated} {input representation}.

Then, we concatenate the rotated data with the original, effectively enriching the input representation. Thus, the augmented feature $\overline{\boldsymbol{s}}_i \in \mathbb{C}^{1 \times M}$ is defined as 
 \begin{align}
\overline{\boldsymbol{s}}_i &=  [\overline{{s}}_{i,1}, \cdots, \overline{{s}}_{i,M} ] \\ \nonumber
 & = [\boldsymbol{S}^{(0,0)}_i, \cdots, \boldsymbol{S}^{(P{-1},Q{-1)}}_i, \hat{\boldsymbol{S}}^{(0,0)}_i, \cdots, \hat{\boldsymbol{S}}^{(P{-1},Q{-1)}}_i  ], 
    \end{align}
{
where $P=Q=32$ denote the dimensions of the spatially aggregated complex input representation, satisfying $PQ=M/2=1024$. The resulting complex coefficients, collected in $\overline{\boldsymbol{s}}_i$, are mapped to the corresponding input apertures through the input-layer formulation in Eq.~(1).

This spatial mapping also provides the physical basis for the phase-domain feature augmentation. Since the original and phase-rotated components are assigned to spatially distinct groups of input apertures, they generally experience different propagation coefficients under the Rayleigh--Sommerfeld diffraction model. Consequently, $\theta_{\rm rot}$ controls their relative phase during coherent propagation and thereby modifies the interference contribution to the received intensity used for classification in Eq.~(5). This mechanism enables the SIM-D$^2$NN to exploit the phase information contained in the complex-valued Level-0 SAR representation more effectively.
}

\subsubsection{Training and Deployment of the SIM-D$^2$NN}
 
{The training of the SIM-D$^2$NN utilizes the physical forward model established in Section II. By applying the Rayleigh-Sommerfeld diffraction principles defined in Eqs. (2)-(5), the network backpropagates the gradients directly to the physical phase shifts $\theta$, as summarized in Algorithm~\ref{alg:Framwork}.}

\begin{itemize}
    \item \textbf{Stage 1 (Offline Training):} 
   This stage aims to determine optimal phase shifts for the metasurface layers. {A subset of data patches is selected from the spatially isolated training set} for offline training, where each patch undergoes complex-valued modulation to enhance robustness to speckle noise and Doppler shifts.  {
Specifically, the channel matrix $\mathbf{H}$ denotes a predetermined channel realization that is kept fixed throughout training and testing.}  These modulated inputs are passed through the SIM-D$^2$NN model, where the learnable phase shifts are updated via backpropagation to minimize the classification loss. {
    To ensure the learned in-wave framework remains resilient to non-deterministic analog environments, we incorporate a stochastic noise-aware training protocol. During each forward pass, i.i.d. additive white Gaussian noise is dynamically injected into the transmission channel. By exposing the network to various random noise realizations across different training epochs, the SIM-D$^2$NN is forced to learn robust features that are invariant to the stochastic thermal fluctuations and hardware noise floor inherent in real-world analog systems.
} This approach allows us to simulate and optimize the behavior of the physical SIM under task-specific constraints.

    \item  \textbf{Stage 2 (SIM-D$^2$NN Deployment):} Once training is complete, the optimized phase shifts are fixed and embedded into the physical metasurface layers. During inference, the incoming raw SAR data are modulated using the same strategy and directly propagated through the SIM{\footnote{
{
In practical deployments, channel acquisition and variations across coherence intervals can be handled by dedicated communication-layer mechanisms, including SIM-specific channel estimation using uplink pilots~\cite{yao2024channel} and channel equalization for temporal variations~\cite{sanchez2022airnn}. Transfer learning can also adapt a SIM model trained on one channel to a new channel. These functions are treated separately from the in-wave inference model considered in this work.}
    }}. The classification decision is made based on the antenna output with the highest received intensity, which corresponds to the predicted terrain class. This feedforward process is entirely in the analog domain, requiring no additional onboard computation.
\end{itemize}
{
This section formulates terrain classification as a constrained phase-optimization problem and introduces a two-stage framework for offline training and online deployment. By addressing the unit-modulus constraints and raw SAR data complexity, the proposed framework bridges digital optimization and analog execution. The next section evaluates its robustness and sensitivity to key operational parameters.
}

\section{Design Evaluation and Parameter Sensitivity}

Following the established framework, this section evaluates how various network configurations, training strategies, and optimization techniques influence the performance of the proposed SIM-D$^2$NN. Ablation experiments are conducted across three classification tasks of increasing complexity: a { binary} classification task within a single geographic zone, and  {three}-class and  {four}-class tasks across diverse regions. Results are summarized in Table~\ref{tab:ablation_all_class_tasks}, where each row represents a specific ablation setting, and each column group reports the precision, recall, F1 score, and overall accuracy for the respective task.

\subsection{Dataset Construction and Setup}

The S1 Level-0 raw SAR data utilized in this work were obtained from the Copernicus Browser \cite{filipponi2019sentinel}. To support a thorough ablation study, separate datasets are constructed for {binary} and multi-class terrain classification. Each dataset is generated using a sliding window approach with a fixed patch size of $128 \times 128$ pixels and a stride of 32 pixels to balance patch diversity and spatial coverage. {
 To eliminate generalization bias from spatial autocorrelation, a block-isolation scheme is enforced along the azimuth dimension by splitting the SAR frame into disjoint training and evaluation swaths. To completely decouple the 75\%  stride overlap, a  288-pixel guard band exceeding twice the patch dimension is introduced at the splitting boundaries, guaranteeing zero pixel-sharing between the two distributions.}  
{
For all three classification tasks, the 10\% sampling setting is applied after block isolation, with approximately 80\% of the selected samples drawn from the spatially isolated training pool and 20\% from the testing pool.
}

\begin{itemize}
    \item \textbf{{Binary} Terrain Classification:}  
    We select patches from a contiguous coastal region near Santos and Sao Paulo, Brazil, containing both land and ocean areas. Manual labeling is conducted using high-quality reference imagery generated through the procedure described in Fig.~\ref{workflow}. The resulting dataset comprises 285,680 patches evenly split between land and ocean categories.

    \item \textbf{{Three}-Class Terrain Classification:}  
    This dataset integrates patches from three geographically distinct regions, each dominated by a specific terrain: desert, forest, or ocean,  
    { where 10\% subset contains 34,436 patches.}

    \item \textbf{{Four}-Class Terrain Classification:}  
    Extending complexity, this dataset includes four terrain categories: desert, ocean, urban, and forest, which are sourced from four geographically separate regions. 
    {The corresponding 10\% subset contains 45,915 patches.}
\end{itemize}
 
In summary, these progressively scalable tasks systematically evaluate the SIM-D$^2$NN framework against increasing intra-class variability and geographic diversity.

 \subsection{Performance Metrics}

{
Given the predicted antenna index $\hat{k}_i = \arg\max_{k \in \mathcal{K}} |y_{i,k}|^2$ from the physical receiver array and the one-hot ground truth label vector $\boldsymbol{o}_i = [o_{i,1}, \dots, o_{i,C}]^T \in \mathbb{R}^{C \times 1}$, let $\hat{c}_i = \arg\max_{c \in \{1,\dots,C\}} o_{i,c}$ denote the ground truth category index for patch $i$. Under our direct in-wave mapping configuration, i.e., $K=C$, the physical antenna index $\hat{k}_i$ directly provides the numerical prediction for evaluation against the target class index $\hat{c}_i$.}
The class-wise counts of True Positives (TP), True Negatives (TN), False Positives (FP), and False Negatives (FN) are computed as:
\begin{equation}
\begin{aligned}
{\rm TP}_{c}\!=\!\sum_{i=1}^{I}\mathbb{I}{(\hat{k}_i = {c}, \, {\hat{c}_i}= {c})}, \;\; {\rm FP}_{c}\!=\!\sum_{i=1}^{I}\mathbb{I}{(\hat{k}_i = {c}, \, {\hat{c}_i} \ne {c})}, \\
{\rm FN}_{c}\!=\!\sum_{i=1}^{I}\mathbb{I}{(\hat{k}_i \ne {c}, \, {\hat{c}_i} = {c})}, \;\; {\rm TN}_{c}\!=\!\sum_{i=1}^{I}\mathbb{I}{(\hat{k}_i \ne {c}, \, {\hat{c}_i} \ne {c})},
\end{aligned}
\end{equation}
where $\mathbb{I}{(\cdot)}$ denotes the indicator function that returns $1$ if the enclosed condition is true, and $0$ otherwise.
Based on these counts, we define the evaluation metrics:
\begin{table*}[!ht]
\centering
\caption{Design evaluation under various terrain classification tasks.}
\renewcommand{\arraystretch}{1.2}
\resizebox{\textwidth}{!}{%
\begin{tabular}{l||cccc|cccc|cccc}
\toprule
\toprule
\multirow{2}{*}{\textbf{Settings}} 
& \multicolumn{4}{c|}{\textbf{{Binary} Terrain Classification}} 
& \multicolumn{4}{c|}{\textbf{{Three}-Class Terrain Classification}} 
& \multicolumn{4}{c}{\textbf{{Four}-Class Terrain Classification}} \\
\cmidrule{2-13}
& $\Delta_{\rm Acc}$ & $\Delta_{\rm Pre}$  &  $\Delta_{\rm Rec}$  & $\Delta_{\rm F1}$
& $\Delta_{\rm Acc}$  &  $\Delta_{\rm Pre}$  & $\Delta_{\rm Rec}$ & $\Delta_{\rm F1}$
& $\Delta_{\rm Acc}$ &  $\Delta_{\rm Pre}$  & $\Delta_{\rm Rec}$ & $\Delta_{\rm F1}$ \\
\midrule
S = 20\%        & 90.35\% & 88.05\%  & 92.65\%  & 90.29\%   & 76.23\% &  77.65\%& 76.67\% & 76.66\% &  66.16\% & 66.74\% & 67.69\%  & 66.70\%  \\
S = 100\%       & 91.56\% & 92.31\%  & 95.17\%  & 93.72\%  & 77.13\%   &79.04\%  & 77.16\% & 78.08\% &  67.14\% & 69.17\%  & 69.15\% & 69.15\% \\
\midrule
$\theta_{\rm rot}$ = 0$^\circ$ & 53.90\%  & 56.30\%  & 87.25\%  & 68.43\%  & 52.40\% & 57.68\%  & 57.89\%  & 56.82\%  &  43.19\%& 45.66\% & 45.44\% &45.29\%  \\
$\theta_{\rm rot}$ = 30$^\circ$     & 69.01\% & 68.05\% & 89.55\% & 77.33\% & 56.26\% & 59.84\%  & 59.76\%  & 59.51\% &  44.95\% & 47.00\%& 46.76\% &  46.58\% \\
$\theta_{\rm rot}$ = 60$^\circ$       & 86.80\% & 85.85\% & 90.61\%  & 88.17\%  &  68.33\% & 66.91\% &69.63\%  & 68.72\%   & 54.14\% &  53.63\%& 54.55\% & 54.56\% \\
\midrule
$\eta = 1 \times 10^{-3}$       & 81.44\% & 83.47\% & 89.53\% & 86.39\% & 62.88\% & 63.68\% & 63.68\% & 63.67\% &  44.24\% &  46.28\% & 46.07\% & 46.11\% \\
$\eta = 1 \times 10^{-4}$        & 63.04\%  & 69.19\% & 83.50\% & 75.66\%  & 46.08\% & 47.43\%  & 47.95\%  & 48.06\% & 31.39\%  &  34.92\%& 34.69\% &  34.71\%\\
\midrule
SGD Optimizer           & 85.22\% & 85.80\% & 90.64\% & 88.16\% & 68.62\% & 67.04\% & 68.94\% & 68.98\% & 52.66\% & 51.81\% &  52.66\%& 52.70\%  \\
RMSProp Optimizer          & 87.66\% &  85.24\% & 89.36\%  & 87.25\% & \textbf{75.43\%} & \textbf{75.35\%} & \textbf{75.24\% }& \textbf{75.29\% }& 62.74\% &  62.20\%& 62.07\% & 62.10\% \\
\midrule
Focal Loss       & 87.66\% & 87.23\%  & 89.61\% & 88.04\% & 75.17\%  & 75.21\% & 75.14\% & 75.17\% & 61.89\%& 62.67\% & 62.69\% & 62.89\%   \\
Label Smoothing  & 87.87\% & 87.75\% & 89.47\% & 88.60\%  & 75.11\% & 75.13\%  & 75.13\% &75.13\%  & 62.24\% &63.87\%  &63.86\%  & 62.86\%\\
\midrule
\textit{\textbf{Proposed Baseline}} & \textbf{\textit{89.51\%}} & \textbf{\textit{88.05\%}} & \textbf{\textit{92.65\%}} & \textbf{\textit{90.29\%}}
                 & \textit{75.27\%} & \textit{75.20\%} & \textit{75.23\%} & \textit{75.21\%}
                 & \textbf{\textit{63.22\%}}  & \textbf{\textit{64.47\%} }& \textbf{\textit{64.31\%}} & \textbf{\textit{64.33\%}} \\
\bottomrule
\bottomrule
 
\end{tabular}
}
\vspace{0.5mm}
\begin{minipage}{\textwidth}
\footnotesize \textit{Note:} The \textbf{\textit{Proposed Baseline }}experiment uses sampling rate S = 10\% data, the phase rotation with $\theta_{\rm rot} = $90$^\circ$, learning rate $\eta = 1 \times 10^{-2}$, AdamW optimizer, and standard CE loss function,
with other rows showing ablations over individual design factors.
\end{minipage}
\label{tab:ablation_all_class_tasks}
\end{table*}
\paragraph{Precision}
Precision measures the proportion of correct positive predictions, indicating prediction reliability and a low false-alarm rate:
\begin{equation}
    \Delta_{\mathrm{Pre}} = \frac{1}{C} \sum_{{c}=1}^{C} \frac{{\rm TP}_{c}}{{\rm TP}_{c} + {\rm FP}_{c}}.
\end{equation}

\paragraph{Recall}
Recall quantifies the model's ability to detect all relevant instances, which is essential in scenarios where missing positive cases are particularly costly:
\begin{equation}
    \Delta_{\mathrm{Rec}} = \frac{1}{C} \sum_{{c}=1}^{C} \frac{{\rm TP}_{c}}{{\rm TP}_{c} + {\rm FN}_{c}}.
\end{equation}

\paragraph{F1 Score}
The F1 score provides a balanced trade-off between Precision and Recall. It is especially useful in the presence of class imbalance or when both false positives and false negatives are critical:
\begin{equation}
    \Delta_{\mathrm{F1}} = \frac{1}{C} \sum_{{c}=1}^{C} \frac{2 \cdot  \Delta_{\mathrm{Pre}_{c}}\cdot  \Delta_{\mathrm{Rec}_{c}}}{ \Delta_{\mathrm{Pre}_{c}} +  \Delta_{\mathrm{Rec}_{c}}}.
\end{equation}
where $\Delta_{\mathrm{Pre}_{c}}= \frac{{\rm TP}_{c}}{{\rm TP}_{c} + {\rm FP}_{c}}$ and $\Delta_{\mathrm{Rec}_{c}}= \frac{{\rm TP}_{c}}{{\rm TP}_{c} + {\rm FN}_{c}}$ denote the precision and recall for the {$c$}-th class, respectively.

\paragraph{Accuracy}
Accuracy provides an intuitive overall measure of correctness across all categories:
\begin{equation}
    \Delta_{\mathrm{Acc}} = \frac{1}{I} \sum_{i=1}^{I} \mathbb{I}({\hat{k}_i = {\hat{c}}_i}).
\end{equation}
Together, these metrics provide a robust and comprehensive basis for evaluating the SIM-D$^2$NN framework across varying classification scenarios and levels of complexity.

\subsection{{Performance Analysis}}
 
\textbf{Sampling Rate ($S$):}  
Table~\ref{tab:ablation_all_class_tasks} illustrates the impact of training data volume by comparing 10\%, 20\%, and 100\% sampling rates. Performance across all tasks improves as the sampling rate increases. In the {binary}  task, $\Delta_{\rm Acc}$ rises from  89.51\%  at  10\%  to 91.56\% with the full dataset. While multi-class scenarios follow this trend, the improvement is smaller. For instance,  {four}-class accuracy increases by only 3.92\%, scaling from 10\% to the full dataset. This confirms that while SIM-D$^2$NN achieves high performance with limited data, further expansion yields diminishing marginal gains in generalization.


\textbf{Phase Rotation ($\theta_{\rm rot}$):}  
The impact of phase rotation is evaluated for $\theta_{\rm rot}=0^\circ$, $30^\circ$, $60^\circ$, and $90^\circ$. The $90^\circ$ setting consistently achieves the best performance across all tasks, yielding $\Delta_{\rm F1}$ gains of approximately 32\% and 42\% over the duplicated $0^\circ$ input in the {binary} and {four}-class tasks, respectively. In particular, the $0^\circ$ setting achieves only 53.90\% accuracy in the {binary} task, while $30^\circ$ and $60^\circ$ provide intermediate improvements.
{
These results are consistent with the phase-domain augmentation mechanism described in Section~III-B. The $0^\circ$ setting corresponds to an in-phase duplicated representation, whereas nonzero $\theta_{\rm rot}$ introduces a relative-phase offset between the two spatially mapped components. Among the tested settings, the $90^\circ$ quadrature configuration provides the most discriminative representation for the considered Level-0 SAR data and SIM-D$^2$NN architecture.
}

        \begin{table*}[!ht]
\centering
\caption{Computational complexity { and performance evaluation} on different models.}
\label{tab:benchmark_complexity}
\scalebox{0.95}{
\begin{tabular}{l|c|c|c|c|c}
\toprule
\toprule
\textbf{Model} & \textbf{Learnable Parameters} & \textbf{FLOPs (Per Sample)} & \textbf{Latency} & \textbf{Energy Efficiency} & \textbf{$\Delta_{\text{Acc}}$} \\

\midrule
Proposed SIM-D$^2$NN (\textit{Phase Only, $\sigma_p = 0$}) & $M \cdot L$ { $= 8,192$} & $\approx 0$ &  {$\sim$ps-ns} &  {Very high}&  {89.51\%} \\
 {Proposed SIM-D$^2$NN (\textit{$\sigma_p = 0.05$})} & { $M \cdot L$ { $= 8,192$}}  & {$\approx 0$} &  {$\sim$ps-ns}&  {Very high}&  {87.58\%$\pm$0.7\%} \\
{Proposed SIM-D$^2$NN (\textit{$\sigma_p = 0.1$})} & { $M \cdot L$ { $= 8,192$}}  & {$\approx 0$} &  {$\sim$ps-ns}&  {Very high}&  { 86.32\%$\pm$1.2\%} \\
\midrule
SIM-D$^2$NN \textit{(Phase \& Amplitude)} & $2M \cdot L$ { $= 16,384$}& $\approx 0$&  {$\sim$ps-ns}&  {Very high}&  {90.70\%} \\
Linear Classifier & $2M \cdot {C}$ { $= 8,192$} & { 32,768}  &  {$\sim\mu$s}&  {Low}&  {92.21\%}\\

MLP & $8{C} \cdot (M+{C})$ { $= 32,800$} & { 131,216}  &  {$\sim\mu$s}&  {Low}&  {93.24\%}\\

{ CNN} & {7,969} & {$7.1 \times 10^5$} & {$\sim$ms}&  {Very low}&  {97.34\%}\\

{ ViT} & {7,872} & {$1.2 \times 10^6$} & {$\sim$ms}&  {Very low}&  {96.26\%}\\
\bottomrule
\bottomrule
\end{tabular}
} 

\vspace{5pt} 
\parbox{0.95\textwidth}{ 
\footnotesize { \textit{Note:}} {Parameters and FLOPs are calculated based on $M=2048$ meta-atoms per layer, number of layers $L=4$, and $C=2$ under the binary task.} {
The reported latency and energy efficiency of the SIM-D$^2$NN refer to the in-wave computation stage rather than the complete end-to-end hardware pipeline.
}
}
\end{table*}
\textbf{Optimizer and Learning Rate ($\eta$):}  
We employ the AdamW optimizer with an initial learning rate of $\eta = 1 \times 10^{-2}$ and a StepLR scheduler that reduces $\eta$ by a factor of 0.1 every 20 epochs. To examine learning-rate sensitivity, we evaluate smaller values of $\eta = 1 \times 10^{-3}$ and $\eta = 1 \times 10^{-4}$. We also assess optimizer sensitivity by replacing AdamW with Stochastic Gradient Descent (SGD) and RMSprop under identical scheduling settings \cite{tieleman2012lecture}.
As shown in Table~\ref{tab:ablation_all_class_tasks}, AdamW with $\eta = 1 \times 10^{-2}$ yields the best performance across all tasks. Lower learning rates, particularly $\eta = 1 \times 10^{-4}$, lead to underfitting and substantially reduce classification accuracy in multi-class settings. In the {four}-class task, $\Delta_{\rm Acc}$ drops from 87.33\% to 64.71\%, with average $\Delta_{\rm Pre}$ and $\Delta_{\rm Rec}$ decreasing by over 20\%. These degradations suggest that insufficient learning rates hinder the formation of discriminative phase mappings for analog computation.
Alternative optimizers such as SGD and RMSprop also underperform, with accuracy dropping by 5\%--10\% relative to AdamW in the {four}-class task. This gap is mainly attributed to the limited effectiveness of non-adaptive optimizers in handling the phase-only parameters of the SIM-D$^2$NN architecture.

\textbf{Loss Function:}  
Beyond the standard CE loss, we also benchmark the focal loss \cite{lin2017focal} and label smoothing function \cite{muller2019does}, which are commonly used to mitigate class imbalance and overconfidence, respectively. 
As shown in Table~\ref{tab:ablation_all_class_tasks}, both loss functions yield performance comparable to the baseline in the {binary} and multi-class settings, with variations in accuracy and F1 score within 1$\sim$2\%. In the more challenging  {four}-class task, focal loss slightly improves recall, while label smoothing provides a marginal gain in precision.  Nevertheless, the standard complex CE loss consistently achieves the best overall performance across all tasks in SIM-D$^2$NN training.

{ The evaluation in this section highlights the critical role of the 90$^\circ$ phase-rotation strategy and model depth in unlocking the learnability of uncompressed raw SAR data. These results confirm that strategic in-wave modulation can mitigate the noise inherent in raw SAR data. These insights establish a robust baseline for the SIM-D$^2$NN, the competitive performance of which is validated through the comparative numerical results in the following section.
}

\section{Numerical Results}
Based on the optimized configurations, this section details the parameter and benchmark settings, followed by a performance analysis of the SIM-D$^2$NN for terrain classification.

\subsection{{Parameter Configurations and }Benchmark Description}

The SIM-D$^2$NN is configured with a metasurface consisting of $M = 2048$ meta-atoms distributed across $L = 4$ layers. The communication system employs a single transmit antenna and $K$ receive antenna elements. {To facilitate direct output mapping, the number of physical receiver antennas is configured to perfectly match the number of target terrain categories, i.e., $K = C$, where $C$ denotes the total number of terrain categories.} Unless otherwise noted, all experiments use a carrier frequency of 5~GHz, a space-to-ground distance of 150~km, half-wavelength element spacing $d_x, d_y = \lambda/2$, and the thickness of the SIM is set to 0.05~m. The model is trained using the {AdamW} optimizer with a CE loss criterion, and the model training is conducted with a batch size of 64, an initial learning rate of 0.01, and 60 epochs. The receiver noise power is fixed at $-104$~dBm{, and the Rician fading factor is set to $\kappa = 20$~dB, where the multi-path component $\mathbf{H}_{\text{NLoS}}$ is assumed to follow a standard Rayleigh fading distribution with each entry distributed as $\mathcal{CN}(0, 1)$ to characterize the LoS-dominant space-to-ground propagation environment}.  
 
We consider the proposed SIM-D$^2$NN with phase-only modulation as the primary model of interest, where each meta-atom applies a fixed-amplitude phase shift,  i.e., $|a_m^l| = 1$. This design is fully analog, hardware-friendly, and serves as the foundation for evaluating metasurface-based analog computation. To benchmark its performance, we compare it against the following reference models: 

\begin{itemize}
    \item \textbf{SIM-D$^2$NN (Phase \& Amplitude):}  
    A relaxed analog variant in which each meta-atom can independently adjust both amplitude and phase, i.e., $a_m^l \in [0, 1]$ and $\theta_m^l \in [0, 2\pi]$. While this configuration increases model expressiveness, it introduces additional hardware complexity and implementation challenges.

    \item \textbf{Linear Classifier:}  
    A digital multilayer perceptron is implemented as a single linear layer without any nonlinear activation functions. This model serves as a baseline to evaluate the expressive power of purely linear transformations in a digital setting, using an architecture consistent in input and output dimensionality with the SIM-D$^2$NN. Specifically, it receives an input vector of dimension $M$ and generates a $K$-dimensional output representing the logits for $K$ terrain classes. Unlike SIM-D$^2$NN, this digital model does not incorporate any layer-wise propagation coefficients or in-wave transformations.
 \begin{figure*}[t]
\begin{center}
    \includegraphics[width=1\textwidth]{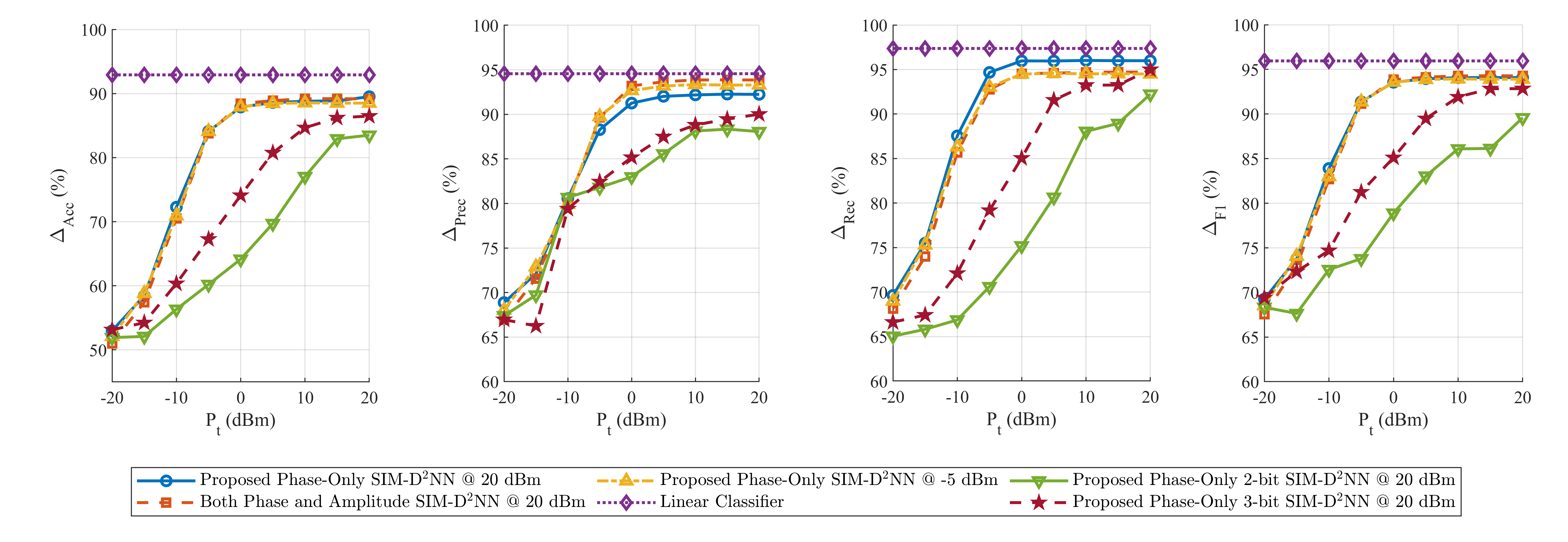}
\end{center}
\caption{ 
  The transmit power P$_t$ versus different performance under the {binary} terrain classification task.}
    \label{fig:Pt}
\end{figure*}
    \item \textbf{Multilayer Perceptron (MLP):}  
    A digital MLP is constructed with two linear layers with a complex-valued ReLU activation. The first layer maps the input from dimension $M$ to $4C$, and the second projects it to a $C$-dimensional output corresponding to the terrain classes. This architecture serves as a strong digital baseline under matched layer dimensions and illustrates the performance benefits introduced by nonlinear computation.



{
\item \textbf{CNN:}
A light CNN is constructed using three convolutional layers. Each convolutional stage employs a $3\times3$ kernel {followed by a complex-valued ReLU activation}, with successive stages progressively reducing the spatial dimensions. The network ends with global average pooling and a fully connected layer, and this structural baseline aligns input configurations with the physical degrees of freedom of the SIM-D$^2$NN.}

{ \item \textbf{Vision Transformer (ViT):}  
A light ViT baseline is constructed using a single-layer, single-head Transformer block. Following the same preprocessing, the input is partitioned into $4 \times 4$ patches, yielding 64 tokens with an embedding dimension of 26. This setup benchmarks digital attention against in-wave diffraction at an equivalent representational scale.}

\end{itemize}


\subsection{Performance Evaluation}
Performance is evaluated across varying transmit power, layer counts, and benchmark models for terrain classification tasks. {
In addition, a benchmark test on the L-band AIRSAR Flevoland dataset is given to further validate the generalization and robustness
of the SIM-D$^2$NN.
}

\subsubsection{Computational complexity and performance evaluation on different models}

Table~\ref{tab:benchmark_complexity} compares the number of learnable parameters, Floating-point Operations (FLOPs), and processing performance across different models. The SIM-D$^2$NN and its variants require $M \cdot L = 8,192$ and $2M \cdot L = 16,384$ parameters, respectively. 
{
Within the in-wave computation stage, the learned feature transformation is realized through passive analog wave propagation without digital arithmetic operations. As such, the corresponding digital FLOPs are negligible.}
{
Physically, the SIM-D$^2$NN leverages passive wave diffraction to execute feature extraction at the speed of light, attaining picosecond-to-nanosecond ($\sim$ps--ns) in-wave computational latency \cite{WC_2024_An_Stacked,TAP_2024_An_Emerging,gao2023programmable}. The passive wave-domain computation has an energy scale on the order of femtojoules (fJ) per operation \cite{gao2023programmable}. These metrics characterize the in-wave computation stage rather than the complete end-to-end processing pipeline.}
{
Furthermore, we evaluate the impact of hardware phase uncertainty by introducing independent phase perturbations drawn from $\mathcal{N}(0,\sigma_p^2)$ at each meta-atom, where $\sigma_p$ denotes the standard deviation in radians. For each $\sigma_p$, the classification accuracy is averaged over 50 independent Monte Carlo trials, with the corresponding standard deviation reported to quantify the statistical uncertainty. The SIM-D$^2$NN achieves accuracies of $87.58\%\pm0.7\%$ and $86.32\%\pm1.2\%$ for $\sigma_p=0.05$ and $0.1$, respectively, indicating relatively stable performance under the considered phase perturbations.
}
\begin{figure*}[t]
  \centering
  \subfigure[SIM-D$^2$NN (No Phase Rotation)]{%
    \includegraphics[width=0.25\textwidth]{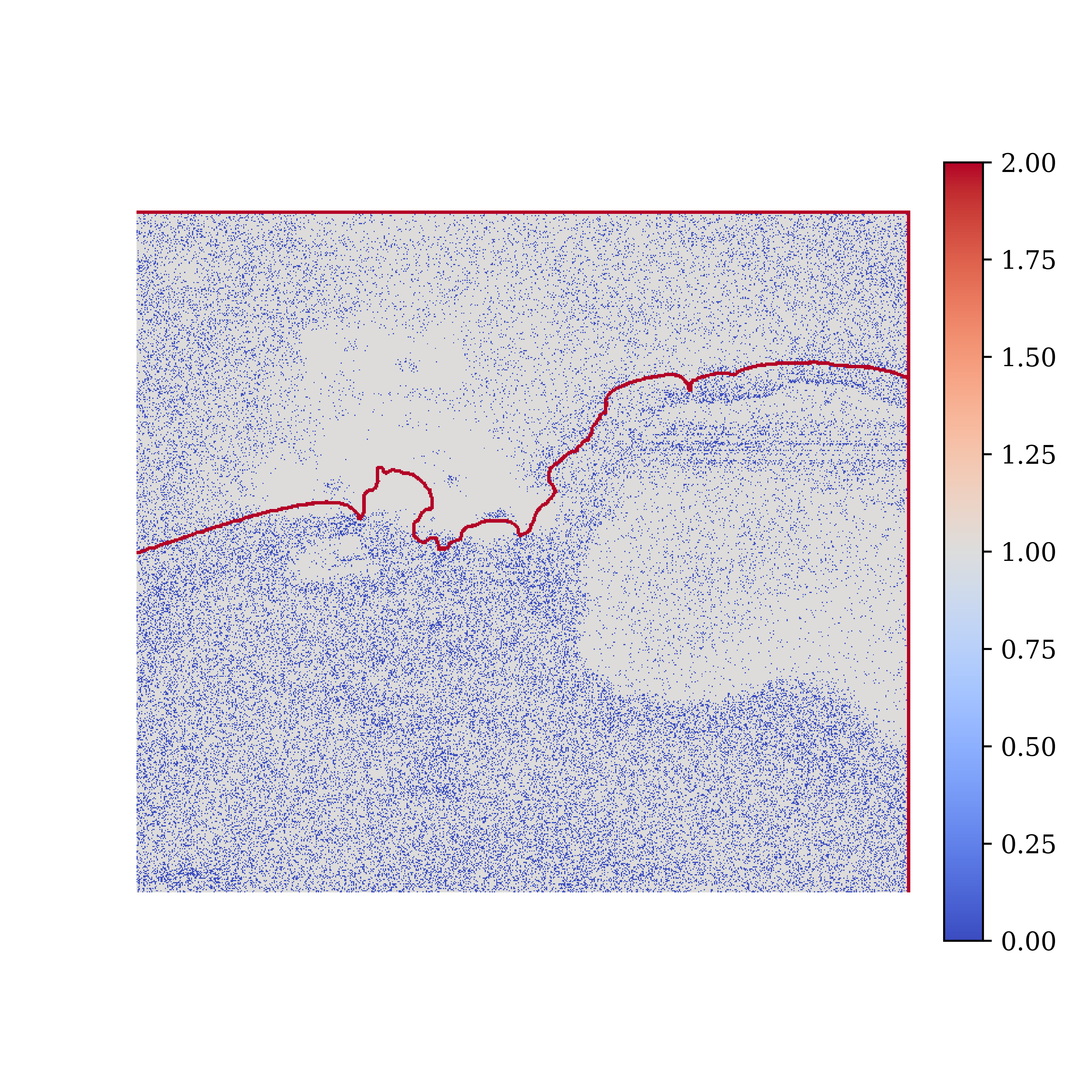}
  }\hspace{0.6em} 
  \subfigure[Proposed SIM-D$^2$NN]{%
    \includegraphics[width=0.25\textwidth]{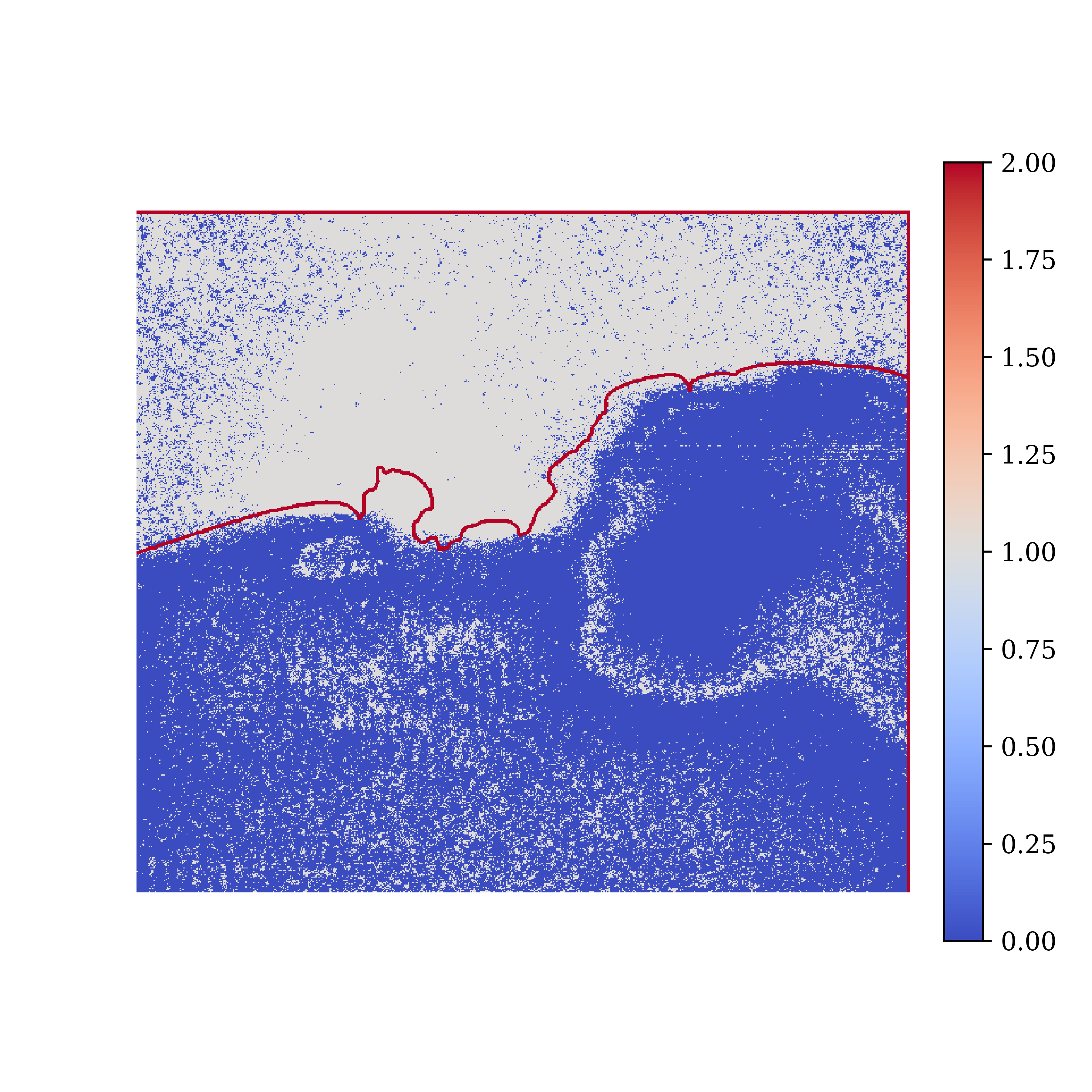}
  }\hspace{0.6em} 
  \subfigure[Linear Classifier]{%
    \includegraphics[width=0.25\textwidth]{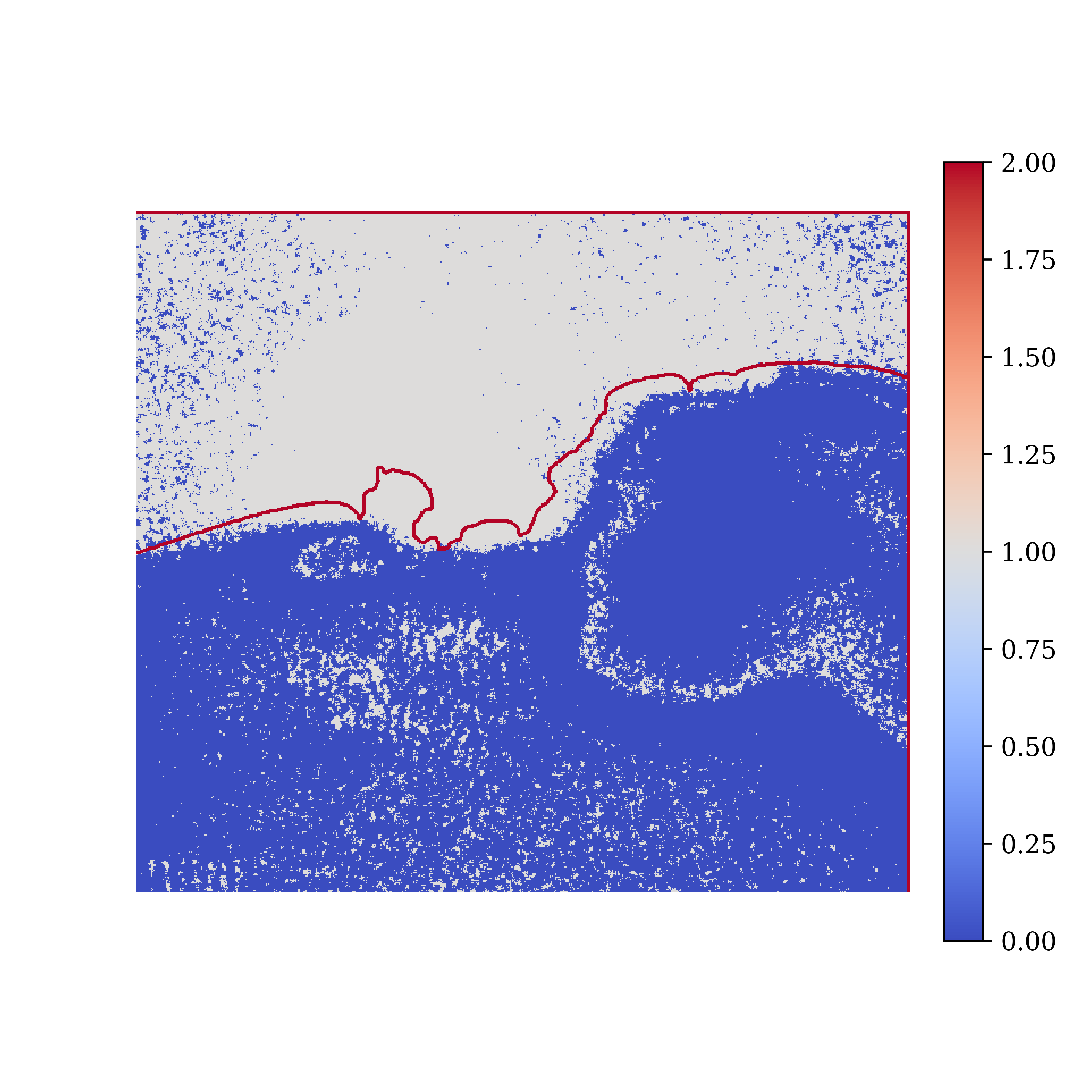}
  }
  \caption{The visualization results on the whole dataset under the {binary} terrain classification task.}
  \label{fig:binary_vis}
\end{figure*}

 In contrast, the digital baseline models entail significant computational overhead. The linear classifier utilizes $2M{C}$ learnable parameters and incurs $8M{C} \approx 3.3 \times 10^4$ FLOPs per sample, accounting for real-imaginary decomposition. The two-layer MLP incurs complexity to $8{C}(M+{C})$ parameters and approximately $1.3 \times 10^5$ FLOPs, where the complex-valued ReLU imposes $8{C}$ FLOPs by separately activating real and imaginary components. 
{
Furthermore, the CNN with intermediate ReLU activations requires approximately $7.1 \times 10^5$ FLOPs per sample and achieves an accuracy of}
{
97.34\%, as reported in Table~\ref{tab:benchmark_complexity}.
To further distinguish the effects of spatial context and nonlinearity, we conduct two additional controlled ablations. Removing the intermediate nonlinear activations reduces the CNN accuracy to 58.21\%, while enlarging the receptive field of the resulting activation-free CNN using dilated convolutions yields a comparable accuracy of 58.47\%.  
These results indicate that the CNN largely benefits from intermediate nonlinear activations, while the SIM-D$^2$NN achieves 89.51\% accuracy with near-zero digital FLOPs through globally connected wave diffraction without intermediate nonlinearities.
}

{
A performance--complexity trade-off is also observed for the ViT baseline. While the ViT achieves a high accuracy of 96.26\%, it requires approximately $1.2 \times 10^6$ FLOPs per sample.}
{
Together with the CNN comparison, these results show that nonlinear digital architectures can achieve higher classification accuracy, whereas the SIM-D$^2$NN substantially reduces the digital computational burden, offering a favorable performance--complexity trade-off for resource-constrained onboard inference.
}

\subsubsection{The transmit power $P_t$ versus four performance metrics on the {binary} terrain classification task}

{
The robustness of SIM-D$^2$NN and the digital linear classifier is evaluated over transmit power $P_t$ from $-$20 to 20~dBm, directly characterizing their resilience to Signal-to-Noise Ratio (SNR) variations as $P_t$ remains linearly proportional to the receive SNR.} As shown in Fig.~\ref{fig:Pt}, increasing $P_t$ consistently improves model performance across all evaluation metrics. For example, the phase-only SIM-D$^2$NN trained at 20~dBm exhibits a substantial improvement in accuracy, with $\Delta_{\rm Acc}$ rising from 52.98\% at $-20$~dBm to 89.51\% at 20~dBm in Fig.~\ref{fig:Pt}(a).
We also investigated the impact of simultaneous phase and amplitude modulation. While this configuration yields marginal gains with 0.19\% at 15~dBm in Fig.~\ref{fig:Pt}(d), doubling learnable parameters reduces cost-effectiveness, justifying phase-only modulation for efficient onboard classification.

To assess the impact of training power, we compare two models trained at $P_t = $20~dBm and $P_t = -$5~dBm, respectively, and evaluate both across the same $P_t$ range. While the trained model at $-$5~dBm achieves slightly better $\Delta_{\rm Pre}$ in the mid-SNR region in Fig.~\ref{fig:Pt}(b), the 20~dBm-trained model yields better overall generalization, particularly in terms of $\Delta_{\rm Rec}$ and $\Delta_{\rm Acc}$ under high-SNR conditions. {Notably, the noise-aware training protocol ensures a stable performance region within the 0--15~dBm range, even when the model is trained at a specific $P_t$.}
{
Meanwhile, to evaluate practical hardware constraints, we assess the quantization gap using 2-bit and 3-bit  phase resolutions. Notably, 3-bit quantization achieves competitive results, reaching 86.48\% accuracy and 92.84\% F1-score at 20 dBm, while 2-bit resolution remains above 83\% accuracy at high SNR in Fig.~\ref{fig:Pt}(a), {proving the architecture's resilience to discrete analog realizations.
 Furthermore, utilizing high-speed PIN diodes with a 2 ns minimum switching time enables nanosecond-level phase refreshment, which is orders of magnitude faster than spaceborne SAR pulse intervals and supports rapid metasurface reconfiguration \cite{qian2025nanosecond}.
}  At last, we benchmark a digital linear classifier with a parameter count comparable to that of SIM-D$^2$NN, which shows that SIM-D$^2$NN approaches digital linear classifier performance at high $P_t$, achieving 92.24\% precision and 95.99\% recall. These results highlight the potential of analog SIM-based computation as a lightweight, power-efficient alternative to digital inference in resource-constrained environments.}

\subsubsection{The visualization results on the whole {binary}  classification dataset}

 Fig.~\ref{fig:binary_vis} presents classification results across the dataset for the { binary}  task using three models: SIM-D$^2$NN without phase rotation, the proposed SIM-D$^2$NN with 90$^\circ$ rotation, and a digital linear classifier. Labels 0, 1, and 2 denote ocean, land, and excluded boundary regions, respectively. Without phase rotation, as shown in Fig.~\ref{fig:binary_vis}(a), the SIM-D$^2$NN yields spatially incoherent predictions resembling random classification, failing to extract meaningful features from the raw SAR data. Conversely, the proposed SIM-D$^2$NN with 90$^\circ$ rotation in Fig.~\ref{fig:binary_vis}(b) generates coherent segmentation with {binary}  boundaries closely aligned with the ground truth.  
 Although minor misclassifications occur near the coastline, these results demonstrate that phase rotation significantly enhances feature separability and spatial consistency without increasing architectural complexity. The digital linear classifier in Fig.~\ref{fig:binary_vis}(c) produces sharp transitions with minor boundary errors, indicating that in-wave inference by SIM-D$^2$NN achieves comparable spatial accuracy under high-SNR conditions.

\subsubsection{The number of layers $L$}
\begin{table*}[!ht]
\centering
\caption{Performance of SIM-D$^{2}$NN at different layer depths under three classification tasks.}
\scalebox{1}{  
\begin{tabular}{c||cccc|cccc|cccc}
\toprule\toprule
\multirow{2}{*}{\textbf{Number of Layers}} 
& \multicolumn{4}{c|}{\textbf{ {Binary} Terrain Classification}} 
& \multicolumn{4}{c|}{\textbf{  {Three}-Class Terrain Classification}} 
& \multicolumn{4}{c}{\textbf{  {Four}-Class Terrain Classification}} \\
\cmidrule{2-13}
& $\Delta_{\mathrm{Acc}}$ & $\Delta_{\mathrm{Pre}}$ & $\Delta_{\mathrm{Rec}}$ & $\Delta_{\mathrm{F1}}$
& $\Delta_{\mathrm{Acc}}$ & $\Delta_{\mathrm{Pre}}$ & $\Delta_{\mathrm{Rec}}$ & $\Delta_{\mathrm{F1}}$
& $\Delta_{\mathrm{Acc}}$ & $\Delta_{\mathrm{Pre}}$ & $\Delta_{\mathrm{Rec}}$ & $\Delta_{\mathrm{F1}}$ \\
\midrule
$L = 1$ & 85.24\% & 86.54\% & 91.86\% & 89.12\% 
    & 65.90\% & 65.44\% & 65.38\% & 65.40\%
    & 50.53\% & 50.42\% & 50.37\% & 50.38\% \\
$L = 4$ & 89.51\% & 88.05\% & 92.65\% &  \textbf{90.29}\%
    & 75.27\% & 75.20\% & 75.23\% & 75.21\%
    & 63.22\% & 64.47\% & 64.31\% & 64.33\% \\
$L = 6$ & \textbf{89.72}\% & \textbf{89.37}\% & \textbf{92.85}\% & 90.11\%
    & \textbf{75.93}\% & \textbf{75.54}\% & \textbf{75.51}\% & \textbf{75.52}\%
    & \textbf{63.61}\% & \textbf{64.54}\% & \textbf{64.51}\% & \textbf{64.52}\% \\
\bottomrule\bottomrule
\end{tabular}
}  
\label{tab:layer_vs_task}
\end{table*}
\begin{figure}[ht]
\setlength{\abovecaptionskip}{0pt}
\setlength{\belowcaptionskip}{0pt}
\centering
\includegraphics[width= 0.5\textwidth]{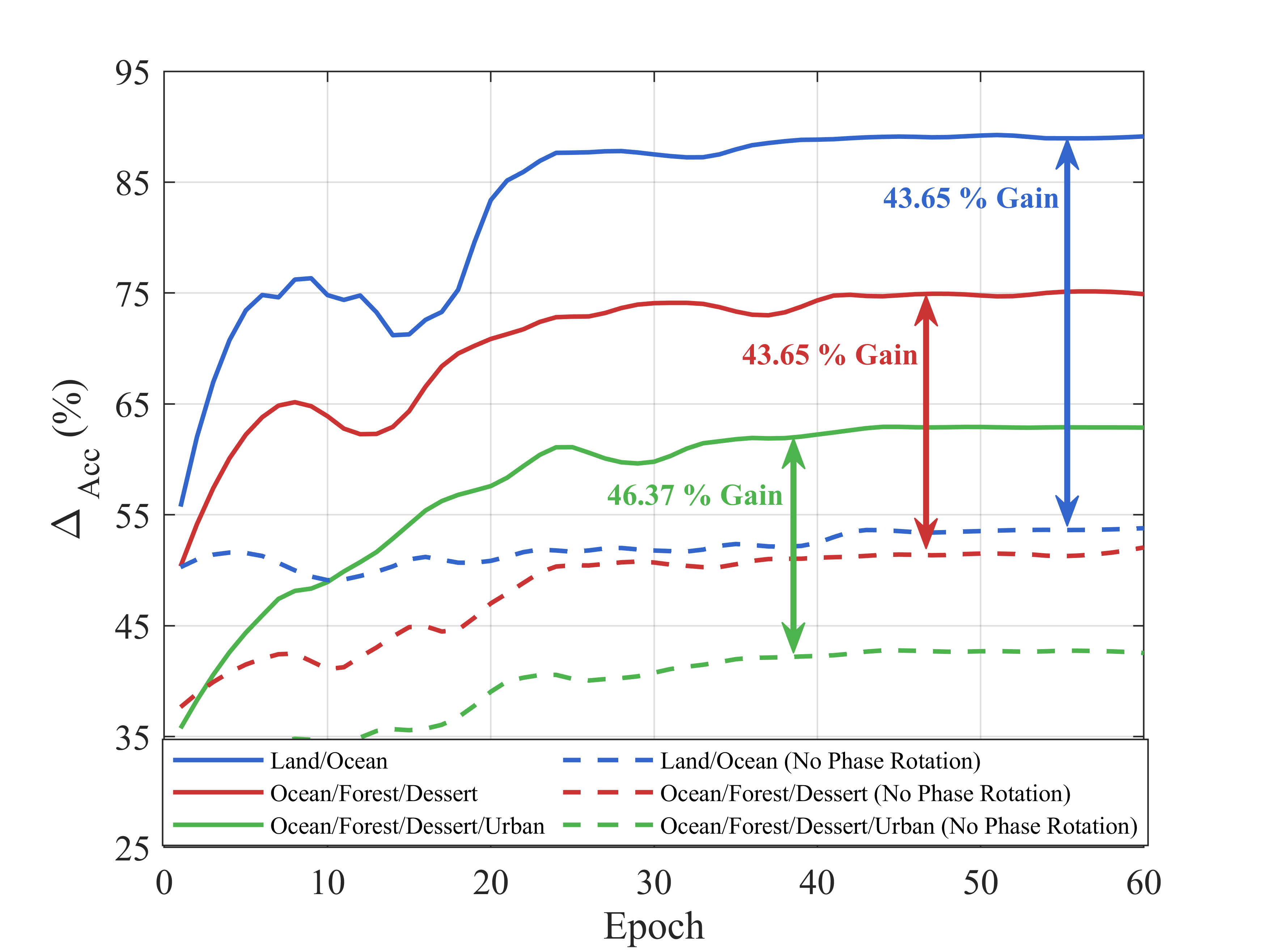}
\DeclareGraphicsExtensions.
\caption{The number of epochs versus testing accuracy.}

\label{fig:convergence}
\end{figure}
We investigate the impact of network depth on the performance of SIM-D$^2$NN across three terrain classification tasks. As shown in Table~\ref{tab:layer_vs_task}, increasing the number of layers from $L = 1$ to $L = 4$ yields substantial improvements across all performance metrics. The gain from increasing depth is especially pronounced in more complex classification tasks. For instance, the relative improvement in $\Delta_{\rm Acc}$ from $L = 1$ to $L = 4$ is 5.0\% for the {binary} classification task, but rises to 14.2\% and 25.1\% for the {three}-class and {four}-class tasks, respectively. Similar trends are observed in $\Delta_{\rm Pre}$, $\Delta_{\rm Rec}$, and $\Delta_{\rm F1}$, all of which exhibit clear gains with increasing depth, particularly in multi-class scenarios.

Nevertheless, the performance gains from $L = 4$ to $L = 6$ are marginal. In the {four}-class task, accuracy increases only slightly from 63.22\% to 63.61\%, and $\Delta_{\rm F1}$ stabilizes around 64.5\%. A similar saturation is observed for $\Delta_{\rm Pre}$ and $\Delta_{\rm Rec}$, which show negligible improvements beyond four layers. This saturation effect suggests that while adding layers initially enhances the model’s representational capacity, deeper configurations offer diminishing returns, particularly as task complexity increases. Therefore, a four-layer SIM-D$^2$NN achieves a favorable balance between performance and hardware complexity, making it a suitable choice for onboard terrain classification across diverse settings.

 \begin{figure}[ht]
\setlength{\abovecaptionskip}{0pt}
\setlength{\belowcaptionskip}{0pt}
\centering
\includegraphics[width= 0.5\textwidth]{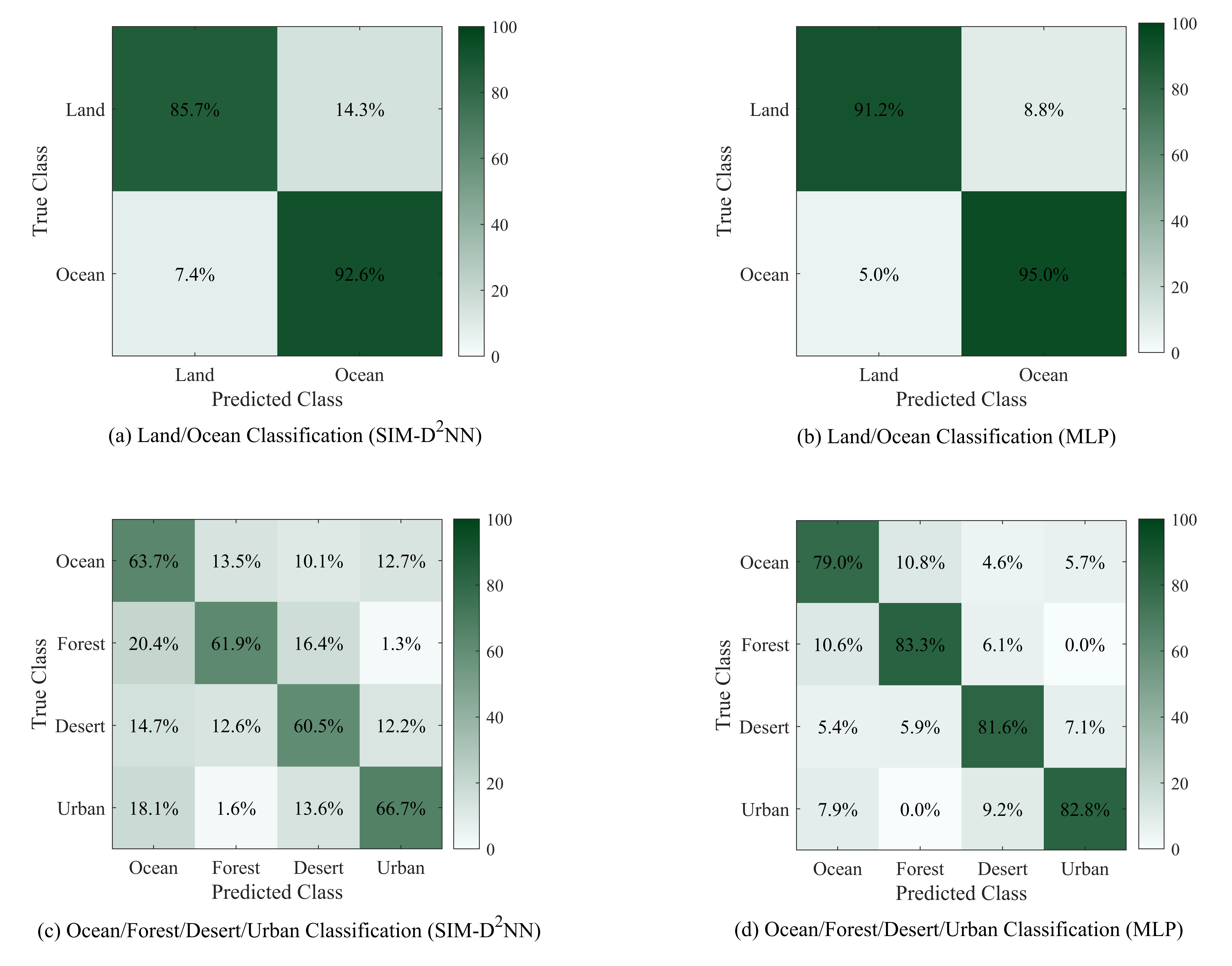}
\DeclareGraphicsExtensions.
\caption{The confusion matrix for both {binary} and  {four}-class terrain classification tasks.}
\label{fig:confusion_matrix}
\end{figure}

\subsubsection{Convergence analysis}
 
Fig.~\ref{fig:convergence} presents the convergence behavior of the SIM-D$^2$NN across three tasks, comparing models trained with and without phase rotation. Models incorporating rotation achieve significantly higher $\Delta_{\rm Acc}$ across all scenarios, with relative accuracy gains of 55.68\% for the {binary}  task, 43.65\% for the  {three}-class task, and 46.37\% for the  {four}-class task. These improvements highlight the critical role of phase rotation in enhancing data diversity and mitigating the impact of speckle noise and phase ambiguity inherent in raw SAR data. While an initial performance drop occurs due to random initialization and unstable analog modulation, the network rapidly learns effective phase patterns, resulting in swift recovery and stable convergence.

{
This robustness is physically rooted in the multi-layer diffractive architecture, which functions as a spatial integrator by enabling each input  {patch} to interfere across the entire metasurface aperture. This mechanism provides inherent physical filtering that mitigates stochastic phase fluctuations induced by speckle and satellite motion jitter. By introducing 90$^\circ$ phase diversity to the complex field, the proposed strategy stabilizes intensity-based detection against coherent interference and platform-induced phase ramps. Consequently, the SIM-D$^2$NN establishes an energy-neutral analog baseline with robust discriminative capacity, even under severe signal-level distortions.}

\subsubsection{Confusion matrix comparison on SIM-D$^2$NN and MLP}

Fig.~\ref{fig:confusion_matrix} compares the confusion matrices of the SIM-D$^2$NN and MLP models for {binary} and  {four}-class terrain classification. In the {binary}  task shown in Fig.~\ref{fig:confusion_matrix}(a) and Fig.~\ref{fig:confusion_matrix}(b), the SIM-D$^2$NN achieves class accuracies of 85.7\% for land and 92.6\% for ocean, closely matching the results of 91.2\% and 95.0\% from the MLP. These results indicate that localized terrain features can be effectively resolved using purely linear in-wave transformations.
Conversely, the limitations of the linear SIM-D$^2$NN are evident in the  {four}-class setting across Fig.~\ref{fig:confusion_matrix}(c) and Fig.~\ref{fig:confusion_matrix}(d). The SIM-D$^2$NN attains accuracies of 61.9\% for forest, 60.5\% for desert, and 66.7\% for urban classes, with 16.4\% of forest patches misclassified as desert. 

In contrast, the MLP maintains over 81.7\% accuracy across all classes, demonstrating superior discriminative capability through its single nonlinear hidden layer. { This accuracy reduction as task complexity scales primarily reflects the inherent difficulty of raw signal processing, which is consistent with Level-0 benchmarks where even optical data yields limited performance \cite{del2025enhancing}. Additionally, the linear nature of wave propagation lacks the nonlinear mechanisms required to resolve highly overlapping features, further limiting discriminative capacity for multi-class tasks.} Even with increased network depth, improvements remain limited as shown in Table~\ref{tab:layer_vs_task}, suggesting that linear stacking is insufficient for complex decision boundaries. These findings underline the importance of incorporating nonlinearity when operating directly on noisy raw SAR data in multi-region scenarios.

    \begin{figure*}[t]
\begin{center}
    \includegraphics[width= 1\textwidth]{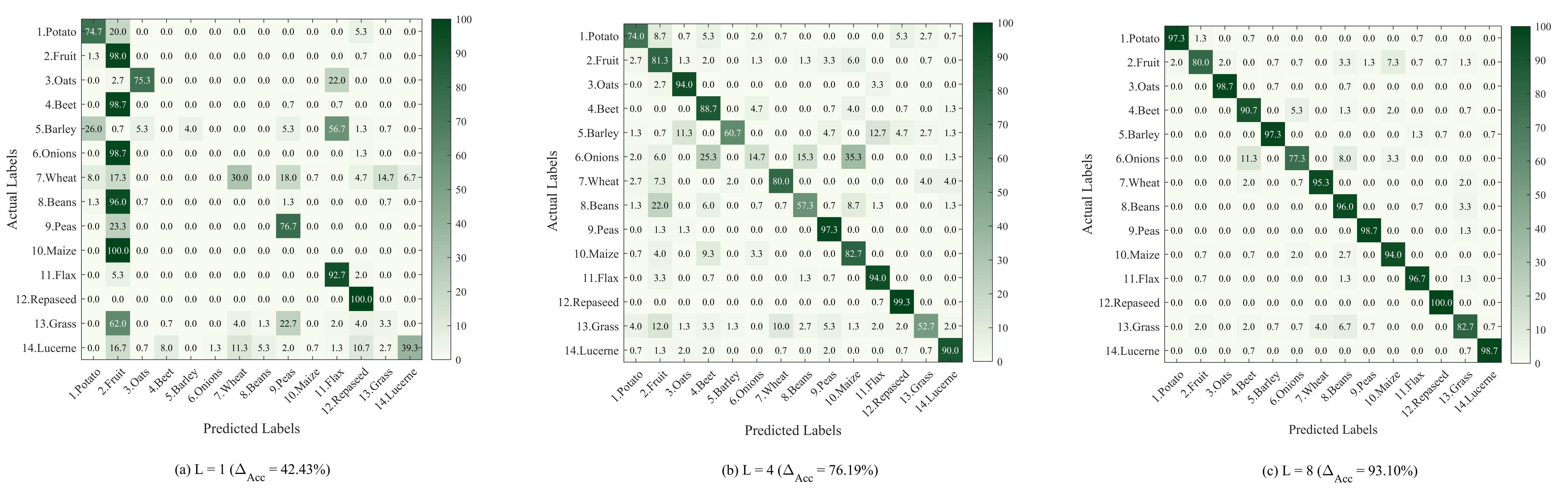}
\end{center}
\caption{ 
 {The confusion matrices under different values of $L$ on the L-band AIRSAR dataset.}}
    \label{fig:confusion_matrix_polsar}
\end{figure*}

\subsubsection{ Benchmark Testing on L-band AIRSAR Flevoland Dataset}

To further {validate the generalization of the SIM-D$^2$NN across sensors and frequency bands, we conduct a benchmark experiment on the L-band AIRSAR Flevoland dataset. Unlike raw SAR data, this preprocessed 14-class dataset provides an upper-bound performance analysis for our analog processing architecture, {where this benchmark is conducted using the linear SIM architecture directly, without employing the phase-rotation augmentation.}} 
Fig.~\ref{fig:confusion_matrix_polsar} presents confusion matrices for varying network depths using the input format from \cite{zhang2017complex}. Classification is performed using received signal intensities, with the metasurface dimensions $M = 864$ and $K = 14$ matched to the input features and target classes. 
As network depth increases, the model exhibits accuracy gains, rising from 42.43\% to 76.19\% and 93.10\% for $L=1, 4, 8$, respectively. These results demonstrate that the SIM-D$^2$NN effectively models fine-grained labels when operating on semantically rich, preprocessed SAR data. This highlights its capacity for linear mapping in clean environments. However, the contrast with the raw SAR results in Fig.~\ref{fig:confusion_matrix} suggests that linear analog networks are highly sensitive to signal quality rather than limited in representational power. This sensitivity underscores the importance of phase-domain augmentation or hybrid designs to enhance robustness when processing uncalibrated raw data in onboard scenarios.

{ The results presented here demonstrate that while the linear SIM-D$^2$NN architecture achieves competitive performance on specific tasks, it reaches operational boundaries when handling complex, multi-region raw SAR data. The comparison with MLPs and high-level SAR products exposes these linear constraints. {Although the square-law non-linearity in final intensity detection provides a foundational physical-domain activation for basic boundaries, such a single-stage terminal non-linearity remains insufficient to map the highly overlapping feature spaces of multi-class raw SAR data without intermediate hidden-layer non-linearities. This characterization of linear limits offers a roadmap for future hybrid or non-linear analog designs discussed in the conclusion.}}

\section{Conclusion}
 
We have proposed a multi-layer SIM-D$^2$NN architecture tailored for onboard terrain classification directly from raw SAR data. To address the high noise and Doppler distortions inherent in such raw signals, a lightweight 90$^\circ$  phase rotation is introduced as an effective data augmentation strategy. By exploiting in-wave analog propagation across multiple programmable metasurface layers, the model optimizes phase shifts under physical constraints, including inter-layer coupling and unit-modulus conditions. Simulation results show that the proposed method achieves a classification accuracy of around 90\% under {binary}  tasks, validating its capacity for high-quality feature mapping in the analog domain.
The SIM-D$^2$NN framework enables efficient onboard inference with minimal reliance on digital processing, substantially reducing data transmission overhead while maintaining competitive performance. 
 

A key future direction is the integration of nonlinear materials and programmable metasurfaces to enable analog nonlinear operations. This would enhance the expressive capacity of SIM-D$^2$NN and allow for more complex decision boundaries required in fine-grained classification. Results from extended terrain classification tasks show that linear analog architectures remain limited when processing noisy raw SAR data.
{Additionally, while our numerical simulations are grounded in established electromagnetic theories,}
{the robustness analysis further considers stochastic phase perturbations, while practical SIM implementations may also experience amplitude variations, parasitic mutual coupling, and temperature-dependent drift. Future research will prioritize hardware-in-the-loop validation using prototypes to characterize these effects, together with fabrication tolerances and material losses, and to develop corresponding calibration strategies.}
{These dual advancements in nonlinear mechanisms and hardware validation will collectively improve the generalization and robustness of analog systems under real-world conditions.}

\bibliographystyle{IEEEtran}
\bibliography{myref}
 
\end{document}